\documentclass[10pt]{article}
\usepackage{amsmath,amsopn,amscd,amsthm,amssymb,xypic,rotating,epic}

\xyoption{all}

\newcommand{\double}       {\baselineskip 14pt}

\newcommand{\HInsert}[2]   {{\immediate\pdfximage width #2  {#1}\pdfrefximage\pdflastximage}}

\newcommand{\cmd}[1]      {\underline{{#1}}}
\newcommand{\cb}          {\begin{tabbing}MMMMM\=MM\=MM\=MM\=MM\=MM\=MM\=MM\=MM\=MM\= \kill}
\newcommand{\ce}          {\end{tabbing}}

\newcommand{\dqt}[1]        {"{#1}"}

\def\emph{\textsl}
\def\em{\sl}
\def\textbf{\pmb}

\def \capstyle{\small}

\newcounter{myremark}
\newcounter{myexample}
\begin{document}

\title{Covid-19 vaccination strategies with limited resources---a model based on social network graphs} 
\author{Simone Santini}
\date{Universidad Aut\'onoma de Madrid}

\maketitle

\double

\begin{abstract}
  We develop a model of infection spread that takes into account the
  existence of a vulnerable group as well as the variability of the
  social relations of individuals. We develop a compartmentalized
  power-law model, with power-law connections between the vulnerable
  and the general population, considering these connections as well as
  the connections among the vulnerable as parameters that we vary in
  our tests.

  We use the model to study a number of vaccination strategies under
  two hypotheses: first, we assume a limited availability of vaccine
  but an infinite vaccination capacity, so that all the available
  doses can be administered in a short time (negligible with respect
  to the evolution of the epidemic). Then we assume a limited
  vaccination capacity, so that the doses are administered in a time
  non-negligible with respect to the evolution of the epidemic.

  We develop optimal strategies for the various social parameters,
  where a strategy consists of (1) the fraction of vaccine that is
  administered to the vulnerable population and (2) the criterion that
  is used to administer it to the general population. In the case of a
  limited vaccination capacity, the fraction (1) is a function of
  time, and we study how to optimize it to obtain a maximal reduction
  in the number of victims.
\end{abstract}

\section{Introduction}
A number of projects for the development of a vaccine for the
SARS-cov2 virus are in advanced stages of development and clinical
trials
\cite{shin:20,escobar:20,mulligan:20,koirala:20,lurie:20,ahmed:20,jeyanathan:20},
and it seems possible that by the first or second trimester of 2021, a
tested, off-the-lab vaccine could be available. However, off-the-lab
availability is not enough: the vaccine will have to be produced and
distributed in great quantities, and this will take time. The dire
reality of world power makes it quite likely that the early production
will the distributed mainly in the richest and most powerful
countries, while many countries will initially receive an amount
insufficient to vaccinate more than a fraction of the population. This
state of affairs poses the problem of designing a vaccination
strategy: who should be vaccinated first so that the limited amount of
vaccine will have a maximum impact, at least while the country is
waiting for more substantial amounts to be delivered? A general
consensus is that medical and paramedical personnel, crucial in an
epidemic, should be the first to receive the vaccine. However,
assuming that once the essential personnel has been vaccinated there
are still doses available, one faces the problem of how to distribute
this surplus among the general population to minimize the number of
casualties while the country waits for more vaccine to arrive.

We develop here an infection spread model suitable for studying
vaccination strategies in the presence of a vulnerable portion of the
population, and we use it to begin a study of vaccination schedules
under various hypotheses regarding the constraints that the limited
availability of vaccine or the limited vaccination capacity may
impose.

The most common models in the study of epidemic diffusion are
derivations of the standard SIR model
\cite{hethcote:00,acemoglu:20,shulgin:98,satsuma:04}, which makes
strong hypotheses about the distribution of the population and its
interactions. In all these models, if $S(t)$ and $I(t)$ are the
numbers of susceptible (healthy) and infected people at time $t$, the
probability that a healthy person enters in contact with an infected
one is proportional to $S(t)I(t)$. Because of this, the model doesn't
account for the different sizes of social circles that different
people may have \cite{barbour:90,britton:07,gonzalez:08}, thereby not
allowing the modeling of targeted vaccination strategies.

We define a richer model of social interaction, one that takes into
account the variability of social circles, and use it to simulate and
evaluate various vaccination strategies. We consider a general model
and ground it using the social and epidemiological data from Spain as
a test-bed%
\footnote{The epidemiological data are derived from the daily reports
  from the \emph{Ministerio de Sanidad}, published in the web page
  www.mscbs.gob.es/en/profesionales/saludPublica/ccayes/alertasActual/nCov/home.htm
}%
. 

The model divides the population in two groups: a more vulnerable
group (conventionally referred to here as the \emph{elderly}) and a
less vulnerable one (the \emph{young}). The parameters for the two
groups have been set considering the Spanish population age 20--64 for
the young, and age $\ge$65 for the elderly. The social connections
among the young have been estimated using average social circle data
\cite{tamarit:18}. The social connections within the group of elderly,
as well as between elderly and young, are parameters that we vary in
our tests.

Using this model, we test various scenarios that differ in the
fraction of vaccine that is administered to the elderly as well as in
the policy that we use to decide which young should be vaccinated
first. We also consider the case in which the vaccination capacity is
limited and the vaccine must be administered over an extended period
of time, not negligible with respect to the spread of the
epidemic. In this case, we use a genetic optimization algorithm to
design an optimal vaccination schedule.

\section{The Model}
Social groups are customarily modeled as random graphs in which nodes
represent individuals and edges represent social interactions between
individuals \cite{barbour:90,newman:01,newman:05}. Let $G=(V,E)$ be
such a graph, with $V=\{1,\ldots,N\}$ and $E\subseteq{V}\times{V}$. We
assume that the graph is undirected without self-loops
($(u,v)\in{E}\Leftrightarrow(v,u)\in{E}$ and $(u,u)\not\in{E}$ for all
$u$, $v$), and let $m=|E|/2$ be the number of edges%
\footnote{The factor 2 is due to the fact that each arc between $u$
  and $v$ is represented twice in $E$, once as $(u,v)$ and once as
  $(v,u)$.}$
$
. The \emph{neighbor} of a node is
\begin{equation}
  N(u) = \bigl\{ v \bigl| (u,v) \in E \bigr\}
\end{equation}
and its \emph{degree} is $d(u) = |N(u)|$. We indicate with
$\bar{d}=\sum_k{d(k)}/N$ the average number of neighbors of nodes in
the graph.  Let $P_k={\mathbb{P}}_u\bigl[d(u)=k\bigr]$ the probability
that a random node, selected with uniform probability, had degree
$k$. One crucial observation in most social networks is the existence
of relatively few people with many contacts, and a \dqt{long tail} of
many people with less and less social contacts
\cite{barabasi:99,wang:11,lazer:09,palla:07}. Specifically, many a
social network exhibit a \emph{power law} or \emph{scale-free}
distribution of the type $P_k\sim{k}^{-\gamma}$ where $\gamma>1$ is an
exponent that determines how \dqt{fat} the log tail is, and that
depends on the specific social network \cite{muchnik:13,arenas:04}.

\bigskip

Several random graph models have been defined to mimic this
behavior. The one that we use here was described in
\cite{barabasi:99,leskovec:08}. The algorithm creates the graph by
adding one node at the time, and is controlled by two parameters, $N$
and $q$, where $N$ is the number of node of the final graph and $q$ a
parameter that determines connectivity. The graph is built in $N$
steps, indexed by a parameter $t=1,\ldots,N$. At step $t$, a new node
is created, node number $t$. At this point the graph already has $t-1$
nodes. Let $d(t,i)$, $i=1,\dots,t-1$ be the number of neighbors of
node $i$ at step $t$. From the newly created node $t$, $q$ edges are
drawn using a \emph{preferential attachment} (or \dqt{rich gets
  richer}) strategy: node $t$ is connected to node $i$ with
probability
\begin{equation}
  \label{pprob}
  P\Bigl[t \leftrightarrow i\Bigr] = 
  \frac{d(t,i)}{\sum_{k=1}^{t-1}d(t,k)}
\end{equation}
The algorithm that generates the graph is shown in
Figure~\ref{algorithm}. The function $\mbox{prefRnd}(V,L)$ generates a
random element of the set $V$ using the list $L=[d(1),\ldots,d(t-1)]$
to determine the probabilities of generating elements of $V$ as in
(\ref{pprob}). It can be shown that this algorithm generates a power
law graph with $P_k\sim{k^{-\gamma}}$, where the exponent $\gamma$
depends on the parameter $q$ \cite{leskovec:08}.
\begin{figure}[thbp]
  \begin{center}
    {\tt
      \cb
      (V, E) = random\_network(n, q) \\
      1.  \> V $\leftarrow$ $\{1\}$ \\
      2.  \> E $\leftarrow$ $\emptyset$ \\
      3.  \> d $\leftarrow$ $[0 | k=1,\ldots, n]$ \\
      4.  \> \cmd{for} k = $2, \ldots \mbox{n}$ \cmd{do} \\
      5.  \> \> V $\leftarrow$ V $\cup \{ \mbox{k} \}$ \\
      6.  \> \> \cmd{for} c = $1, \ldots \mbox{q}$ \cmd{do} \\
      7.  \> \> \> u $\leftarrow$ prefRnd($d[1],\ldots, d[k-1]$) \\
      8.  \> \> \> E $\leftarrow$ E $\cup \{(\mbox{k}, \mbox{u})\}$ \\
      9.  \> \> \> d$[\mbox{u}]$ $\leftarrow$ d$[\mbox{u}]$ + 1 \\
      10. \> \> \cmd{od} \\
      11. \> \> d$[\mbox{k}]$ $\leftarrow$ m \\
      12. \> \cmd{od} \\
      13. \> \cmd{return} (V,E)
      \ce
    }
  \end{center}
  \caption{\capstyle The algorithm that generates the preferential
    attachment graph. The value $d[k]$ keeps track of the number of
    neighbors of node $k$ at each stage of the construction. The
    function $\mbox{prefRnd}(V,L)$ picks a random element of $V$ with
    the probabilities given by the elements of $L$ via (\ref{pprob}).}
  \label{algorithm}
\end{figure}

The complete model is composed of two such graphs interconnected
through preferential attachment. First, two power-law
graphs are generated. The first is a graph $G_y=(V_y,E_y)$ of young
people generated with parameters $N_y$ and $q_y$ ($|V_y|=N_y$),
resulting in a graph with exponent $\gamma_y$ and $\bar{d}_y$ average
neighbors per node. The second graph, of elderly people,
$G_e=(V_e,E_e)$ is generated with parameter $N_e$ and $q_e$
($|V_e|=N_e$), resulting in a graph with exponent $\gamma_e$ and
$\bar{d}_e$ average neighbors per node.

Finally, the two graphs are connected: a number $\bar{d}_{ey}$ is
selected, which will be the average number of young relations that an
elderly person has. Then $N_e\cdot\bar{d}_{ey}$ edges are generated
between the two graphs, selecting an elderly person at random with
uniform probability, a young person using preferential attachment, and
connecting them. The algorithm that joins the two graphs is shown in
Figure~\ref{joinalg}
The rationale for using preferential attachment here is that certain
young people (e.g., social workers, cleaning personnel, delivery
people, etc.) have contact with many elders, while the majority of
young people have little contact with them. The procedure allows us to
create a model in which the interrelations within the groups and the
relations between groups can be independently controlled.
\begin{figure}
  \begin{center}
    {\tt
      \cb
      ($V$, $E$) = join($G_y=(V_y,E_Y)$, $G_e=(V_e,E_e)$, $\bar{d}_{ey}$) \\
      1.  \> $V$ $\leftarrow$ $V_y \cup V_e$ \\
      2.  \> $E$ $\leftarrow$ $E_y \cup E_e$ \\
      3.  \> $N$ $\leftarrow$ $|V_e|\cdot\bar{d}_{ey}$ \\
      4.  \> $d$ $\leftarrow$ $[1 | n \in V_y]$ \\
      5.  \> \cmd{for} $k \leftarrow 1, \ldots, N$ \cmd{do} \\
      6.  \> \> $u$ $\leftarrow$ unif($V_y$) \\
      7.  \> \> $v$ $\leftarrow$ prefRnd($V_y$, $d$) \\
      8.  \> \> $d[v]$ $\leftarrow$ $d[v]$ + 1 \\      
      9.  \> \> $E$ $\leftarrow$ $E \cup \{(u,v), (v,u)\}$ \\
      10. \> \cmd{od} \\
      11. \> \cmd{return}($V$, $E$)
      \ce
    }
  \end{center}
  \caption{Algorithm that joins two graphs through preferential
    attachment. It assumes that $V_y\cap{V_e}=\emptyset$, that is, the
    identifiers of the nodes for the elderly and those for the young
    are different, Both $V_y$ and $V_e$ are ordered list, allowing to
    maintain the correspondence between the nodes and the values in
    $d$. The function $\mbox{unif}(V)$ returns and element of $V$
    chosen randomly with uniform probability.}
  \label{joinalg}
\end{figure}

\subsection{Infection dynamics}
Each node $u$ of the graph represents an individual which, at any
given time $t$, can be in one of four states, stored in the parameter
$\mbox{state}_t[u]$: \emph{susceptible} ($S$), \emph{infected} ($I$),
\emph{recovered} ($R$), or \emph{victim} ($V$). People transition from
one state to another depending on their contacts and their recovery
probability according to the following diagram:
\begin{equation}
  \label{infdyn}
  \xymatrix{
    *++[o][F]{S} \ar[d]_{\beta} \\
    *++[o][F]{I} \ar[d]_{\rho} \ar[r]^{\phi} &  *++[o][F]{R} \ar[ul]_{\epsilon} \\
    *++[o][F]{V}
  }
\end{equation}
The only transition that depends on the interaction between nodes is
$S\rightarrow{I}$, whereby a susceptible individual becomes
infected. The probability of this transition depends on the parameter
$\beta$ as well as on the number of infected neighbors of a given
individual. The model assumes that an individual interacts with their neighbors uniformly%
\footnote{This is the strongest assumption that we make in out
  model. In reality, of course, each one of us interacts with some of
  our social relations more often than with others. The problem can be
  fixed by transforming the graph into a multi-graph, in which the
  number of edges between two nodes are proportional to the density of
  contacts between them. We currently have no good model to indicate
  how these multiple edges should be created.}%
. So, given a node $u$ in state $\mbox{state}_{t-1}[u]=S$ at time
$t-1$, a node $v\in{N(u)}$ is picked randomly with uniform
probability. If $\mbox{state}_{t-1}[v]=I$ then, with probability
$\beta$ the node goes to $\mbox{state}_{t}[u]=I$ and, with probability
$1-\beta$, it remains in $\mbox{state}_{t}[u]=S$.

The other states evolve independently of the state of the neighbors of
$u$, and do so according to the schema of Table~\ref{transition}.
\begin{table}
  \begin{center}
    {\tt
      \begin{tabular}{|c|c|c|}
        \hline
        $\mbox{state}_{t-1}[u]$ &     $\mbox{state}_{t}[u]$ & with probability \\ 
        \hline
               &  $R$  &  $\phi$ \\
        \cline{2-3}
        $I$    & $V$   &  $\rho$ \\
        \cline{2-3}
               & $I$   & $1-\phi-\rho$ \\
        \hline
        $R$    & $S$   & $\epsilon$ \\
               & $R$   & $1-\epsilon$ \\
        \hline
        $V$    & $V$  &  $1$ \\
        \hline
      \end{tabular}
    }
  \end{center}
  \caption{\capstyle Probabilities of transition for a node from the
    states $I$, $R$, $V$. Transition from state $S$ to $I$ depends on
    the contact with the neighbors, and it is not included in this
    schema (see the text for that transition).}
  \label{transition}
\end{table}
These transitions refer to non-vaccinated individuals. A separate
indicator for each individual indicates whether the individual was
vaccinated.  Individuals can be vaccinated only if they are in state
$S$ (viz., only healthy individuals are vaccinated), and once
vaccinated they will permanently stay in state $S$ (viz., the vaccine
is 100\% effective).

\section{Parameters selection}
The model depends on two classes of parameters. One the one hand,
\emph{social} parameters ($N_y$, $N_e$, $\bar{d}_{ey}$; $q_y$ and
$q_e$, which determine $\bar{d}_y$ and $\bar{d}_e$, respectively)
determine the structure of the graph, while \emph{epidemiological}
parameters (those of (\ref{infdyn}) for young and elderly, that is,
$\beta_y$, $\phi_y$, $\rho_y$, $\epsilon_y$, $\beta_e$, $\phi_e$,
$\rho_e$, $\epsilon_e$) determine the spread of the infection in each
neighborhood of the graph. The epidemic parameters are not observable,
but they can be derived based on observable values, as shown in
Section \ref{epidesec}. The observables used here are relative to the
early spread of covid-19 in Spain, specifically the period up to March
25, when the effects of the lock-down decreed by the government on
March 14 began to have a measurable effect effect on the spread of the
epidemic.

\subsection{Social parameters}
\label{social}
The tests use graphs with $N=5,000$ nodes, with 80\% of the population
being young ($N_y=4,000$) and 20\% being elderly ($N_e=1,000$). We
executed preliminary tests with various values of $N$, observing that
for $N>1,000$ the results were stable and their qualitative
characteristics did not change.  The social graph of the young is
fixed, generated with $q=20$, resulting in $\gamma_y=1.77$ and
$\bar{d}_y=16.9$. These values were chosen to be in compliance with
the model in \cite{tamarit:18}. Then, four different graphs are
generated, with different values of the social connection between the
elderly ($\bar{d}_e$) and of the connection between young and elderly
($\bar{d}_{ey}$)%
\footnote{The actual tests were carried out with 16 graphs with
  different combinations of these parameters. For the sake of
  presentation, we show only four graphs, which exemplify the
  different types of results that we obtained.}%
. These data result in four graphs, identified by the codes in
Figure~\ref{codes}.
\begin{table}
  \begin{center}
    {\tt
      \begin{tabular}{|c||c|c|}
        \hline
                      & \multicolumn{2}{|c|}{$\bar{d}_{ey}$} \\
        \cline{2-3}
        $\bar{d}_{e}$ &  $0.5$    &   $3.0$     \\
        \hline
        \hline
        $1.0$         & A$\alpha$ &   A$\beta$   \\
        \hline
        $7.7$         & B$\alpha$ &   B$\beta$   \\
        \hline
      \end{tabular}
    }
  \end{center}
  \caption{\capstyle Parameters and codes of the four different graphs we use
    to analyze the spread of the infection.}
  \label{codes}
\end{table}
Graphs of type \textbf{A} are characterized by sparse relations between
the elderly, that is, each elderly person interacts with few other
elderly people, while in graph of type \textbf{B} there are many more
elderly-to-elderly interactions. Graphs of type \textbf{$\alpha$} have few
cross-age interactions: each elderly person is in contact, on average,
with only a few young people, while \textbf{$\beta$} graphs have denser
interaction. These distinctions will allow us to analyze the effects
of different vaccination strategies when used in conjunction with other
sanitary measures such as the (relative) isolation of categories at
risk or the reduction of contacts between these categories and the
general population.

\subsection{Epidemiological Parameters}
\label{epidesec}
In order to determine the hidden epidemiological parameters $\phi$,
$\rho$, and $\epsilon$ from observable data, we use an SIR model as a
first approximation. At this time, very few cases of reinfection have
been reported \cite{alizargar:20,kirkcaldy:20} so, for the time being,
one can assume that the immunity is permanent (at least in the time
span considered in the simulation), and set $\epsilon_y=\epsilon_e=0$.

Assume a population of $I(t)$ infected people maintained in
isolation. In the one-compartment SIR model, their evolution is
described by the equations
\begin{equation}
  \label{booh}
  \begin{aligned}
    \frac{d I}{d t} &= -\phi I - \rho I \\
    \frac{d R}{d T} &= \rho I \\
    \frac{d V}{d T} &= \phi I 
  \end{aligned}
\end{equation}
The solution for $I(t)$ is given by 
\begin{equation}
  \label{Isolve}
  I(t) = I_0 \exp\bigl(-(\phi+\rho)t\bigr)
\end{equation}
where $I_0$ is the initial number of infected individuals.  The
average time a person stays infected before they either die or recover
is $1/(\phi+\rho)$. Assuming the day as the time unit, and an average
$D$ days duration of the disease, independently of age, one obtains:
\begin{equation}
  \label{psum}
  \phi_y + \rho_y = \phi_e + \rho_e = \frac{1}{D}
\end{equation}
The values of the individual parameters are determined using the
measured letality, that is, the number of victims divided by the
number of cases. Let $L_y$ and $L_o$ be the letality for young and old
people, respectively. From (\ref{Isolve}) and the first of
(\ref{booh}) we have
\begin{equation}
  V(t) = \phi\int_0^tI(t)\,dt = \frac{\phi}{\phi+\rho}I_0\Bigl[1 - \exp\bigl(-(\phi+\rho)t\bigr)\Bigr]
\end{equation}
For large $t$, 
\begin{equation}
  V(t) \approx \frac{\phi}{\phi+\rho}I_0
\end{equation}
Therefore the letality in terms of the $\phi$s and the $\rho$s can
be expressed as
\begin{equation}
  \begin{aligned}
    \frac{\phi_y}{\phi_y+\rho_y} &= L_y \\
    \frac{\phi_e}{\phi_e+\rho_e} &= L_e
  \end{aligned}
\end{equation}
These equations, together with (\ref{booh}) yield 
\begin{equation}
  \begin{aligned}
    \phi_y &= \frac{L_y}{D} \\
    \phi_e &= \frac{L_e}{D} \\
    \rho_y &= \frac{1}{D}(1-L_y) \\
    \rho_e &= \frac{1}{D}(1-L_e) 
  \end{aligned}
\end{equation}
The value of $\beta$ is derived from its relation to the unitary
infection factor $R_0$, which is given by
\begin{equation}
  R_0 = \frac{\beta}{\phi+\gamma}
\end{equation}
resulting in 
\begin{equation}
  \begin{aligned}
    \beta_y = \frac{R_{0,y}}{D} \\
    \beta_e = \frac{R_{0,e}}{D} 
  \end{aligned}
\end{equation}
The values used in the simulations of Section~\ref{methods} are given
in Table~\ref{epivals}.
\begin{table}
  \begin{center}
    \begin{tabular}{|l|l|l|l|}
      \hline
      \multicolumn{2}{|c|}{Observable} &\multicolumn{2}{|c|}{Model} \\
      \hline
      Parameter & Value &  Parameter & Value \\
      \hline 
      $D$       & 18 days & $\beta_y$ & 0.1333 \\
      $R_{0,y}$  & 2.4 & $\beta_e$ & 0.2667 \\
      $R_{0,e}$  & 4.8 & $\phi_y$  & 0.0011 \\
      $L_y$     & 0.02 & $\phi_e$  & 0.0111 \\
      $L_e$     & 0.2  & $\rho_y$  & 0.0544 \\
      \cline{1-2}
      \multicolumn{2}{c|}{} & $\rho_e$  & 0.044 \\
      \cline{3-4}
    \end{tabular}
  \end{center}
  \caption{\capstyle Parameter values used in our simulation. The
    observable parameters in the first part are derived from the
    epidemic data for Spain in the period prior to March 30, before
    the lock-down showed its effects. The unobservable parameters in
    the second part are derived from those, and are those used in the
    model.}
  \label{epivals}
\end{table}

\section{Methods}
\label{methods}
In the first set of simulations, we assume that all vaccination is
done on a healthy population before the onset of the epidemic. In the
case of an ongoing epidemic such as the one of covid-19, this
corresponds to a scenario in which only the still healthy population
is considered, keeping it effectively isolated from the people who
were already infected. We simulate the different vaccination scenarios
given by the parameters specified below and then, once the suitable
number of people has been vaccinated, we simulate the unfolding of the
epidemic, keeping track of the victims as a function of time. The
simulations are characterized by three parameters:

\begin{description}
\item[\textbf{F}:] Total available vaccine during the period of the
  simulation, expressed as a fraction of the healthy population that
  can be vaccinated. We are especially interested in modeling
  scenarios of scarcity of vaccine. The values used in the simulations
  are $F=0.05,0.1,0.2$.
\item[\textbf{$\psi$}:] The fraction of the available doses (viz., the
  fraction of $F$) that is administered to the elderly. The
  simulations use the values $\psi=0,0.25,0.5,0.75,1$, where $\psi=0$
  means that only the young are vaccinated, while $\psi=1$ means that
  only the elderly are vaccinated. The parameters are chosen so that
  $F\cdot{\psi}$ is in any case no greater than the fraction of
  elderly people in the population. If $F\cdot{\psi}$ is greater than
  this value then, once all the elderly have been vaccinated, the
  notion of \dqt{vaccination strategy} would cease to be useful: all
  the vaccine that is left after vaccinating all the elderly would be
  administered to the young.
\item[\textbf{mode}:] The criterion with which the vaccine is
  administered to the young. The fraction of the vaccine allotted to
  the elderly is always distributed uniformly inside the group. The
  fraction that is used to vaccinate the young is allocated according
  to one of the following modes:
  \begin{description}
  \item[\textbf{popular}:] the vaccine is administered in order of
    \dqt{popularity}, that is, the nodes in the graphs with the most
    neighbors are vaccinated first;
  \item[\textbf{connected}:] the vaccine is administered in order of
    connections with the elderly: the young people with the most
    connections to the elderly are vaccinated first.
  \end{description}
  The rationale for the popular mode is that by vaccinating the people
  with the most connections will slow down the spread of the infection
  by removing dangerous \dqt{hubs}%
  \footnote{It is know that the patient zero in parts of Northern
    Italy was a person that had traveled to China and that,
    unfortunately, was very popular in his hometown. Meeting with many
    friends and family in the short time before the onset of symptoms
    contributed to the rapid initial spread of the virus.}%
  . The rationale for the connected mode is to protect the
  vulnerable population by vaccinating the people that can spread the
  infection from the highly socialized young people (where the spread
  is rapid) to the less socialized but vulnerable elderly.
\end{description}

The duration of the simulations is fixed and is set to $T=100$ days.
\begin{figure*}[thbp]
  \begin{center}
    \begin{tabular}{cccc}
      \multicolumn{4}{c}{Mode: \textbf{popular}; vaccination level: \textbf{5\%} ($F=0.05$)} \\
      \hspace{-2em}\HInsert{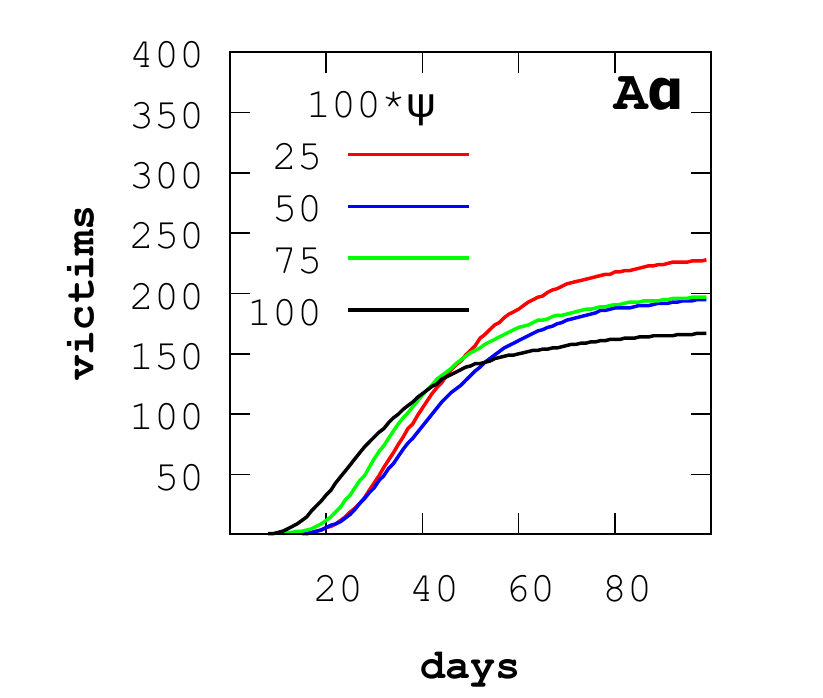}{0.35\textwidth} &
      \hspace{-5.5em}\HInsert{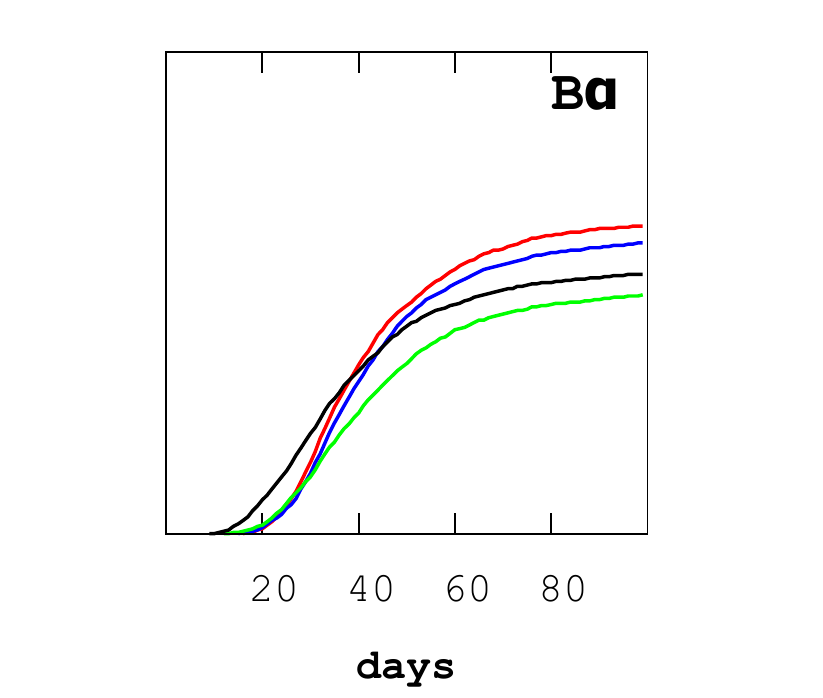}{0.35\textwidth} &
      \hspace{-6.5em}\HInsert{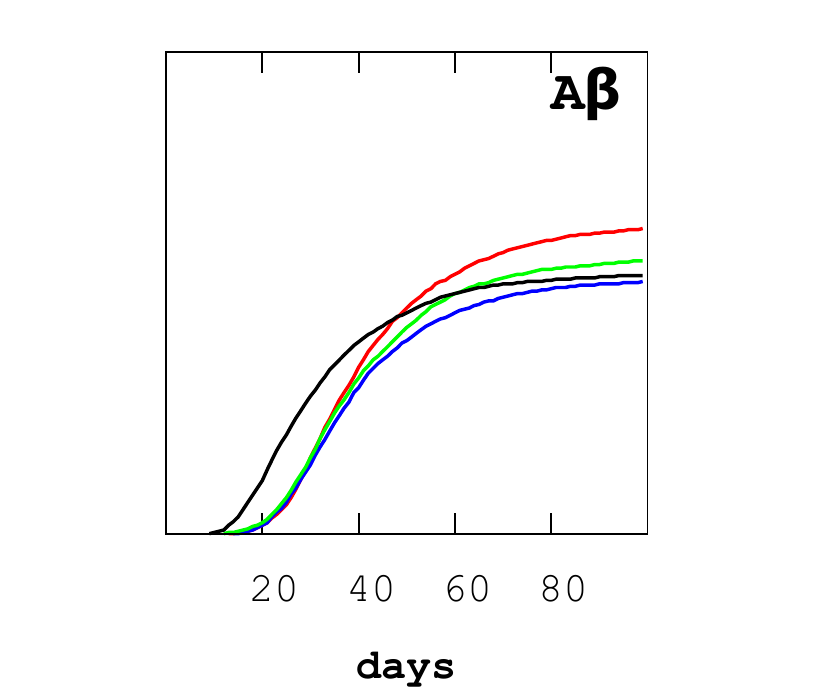}{0.35\textwidth} &
      \hspace{-5.3em}\HInsert{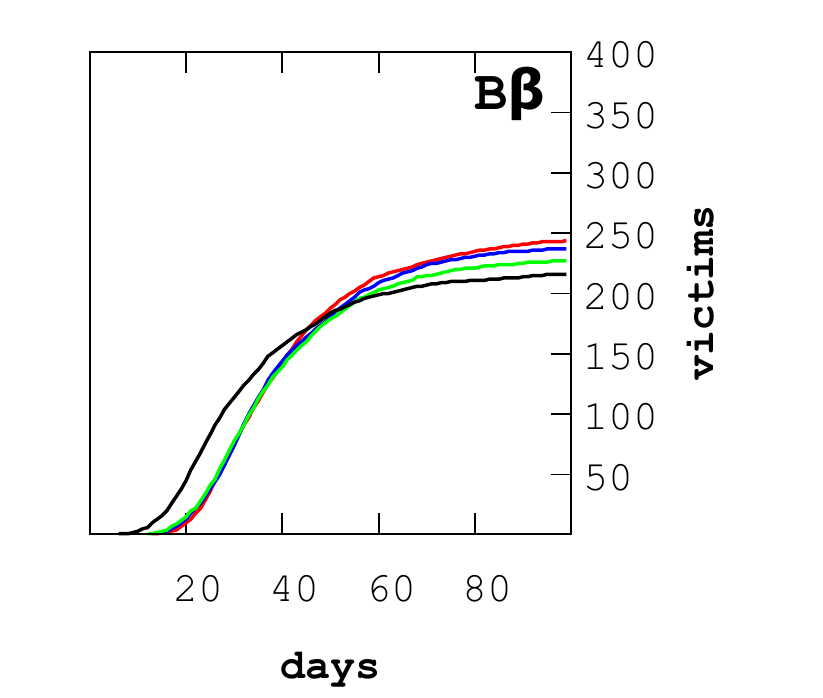}{0.35\textwidth} \\
      \hspace{-1em}(a) & 
      \hspace{-6em}(b) & 
      \hspace{-7em}(c) & 
      \hspace{-9em}(d) 
    \end{tabular}

    \vspace{-1em}

  \end{center}
  \caption{\capstyle Results for the infinite delivery capacity
    vaccination. Each curve is drawn for a given fraction of the
    vaccine administered to the elderly people ($0.25,0.5,0.75,1$) and
    tracks the number of victims during the $100$ days of the
    simulation. Here, $F=0.05$ ($5\%$ of the population are
    vaccinated) and the mode is \textbf{popular}.  }
  \label{pop_05}
\end{figure*}
\begin{figure*}[thbp]
  \begin{center}
    \begin{tabular}{cccc}
      \multicolumn{4}{c}{Mode: \textbf{connected}; vaccination level: \textbf{5\%} ($F=0.05$)} \\
      \hspace{-2em}\HInsert{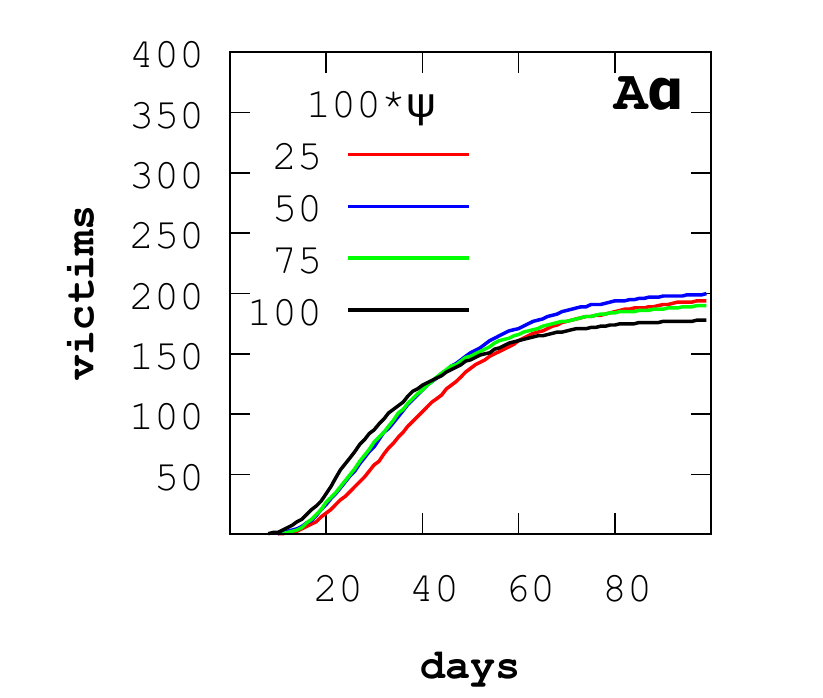}{0.35\textwidth} &
      \hspace{-5.5em}\HInsert{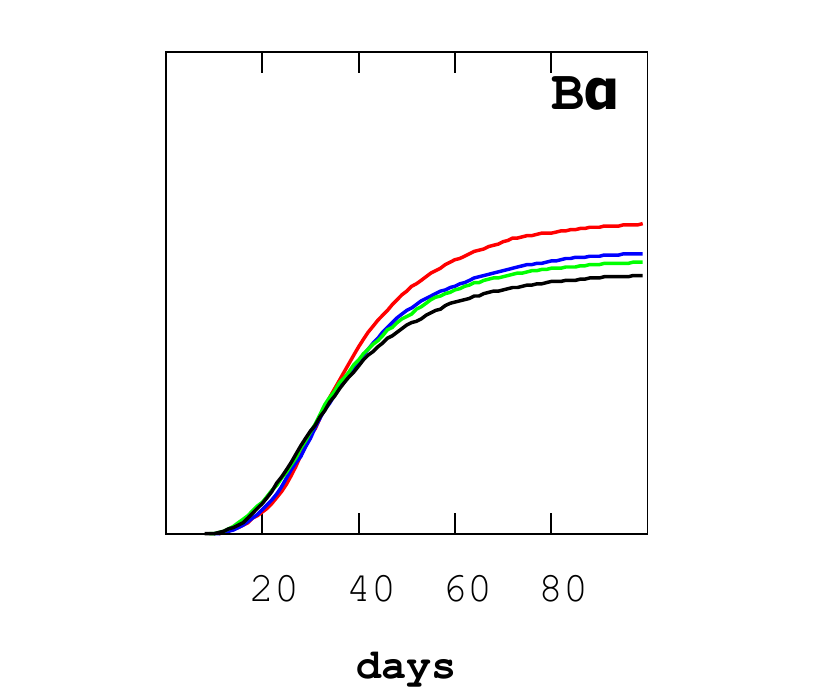}{0.35\textwidth} &
      \hspace{-6.5em}\HInsert{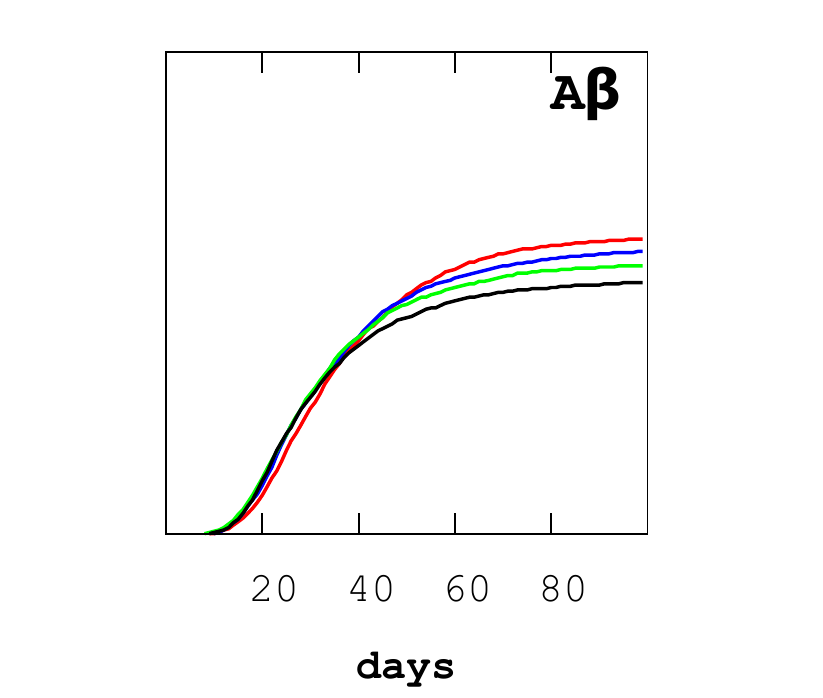}{0.35\textwidth} &
      \hspace{-5.3em}\HInsert{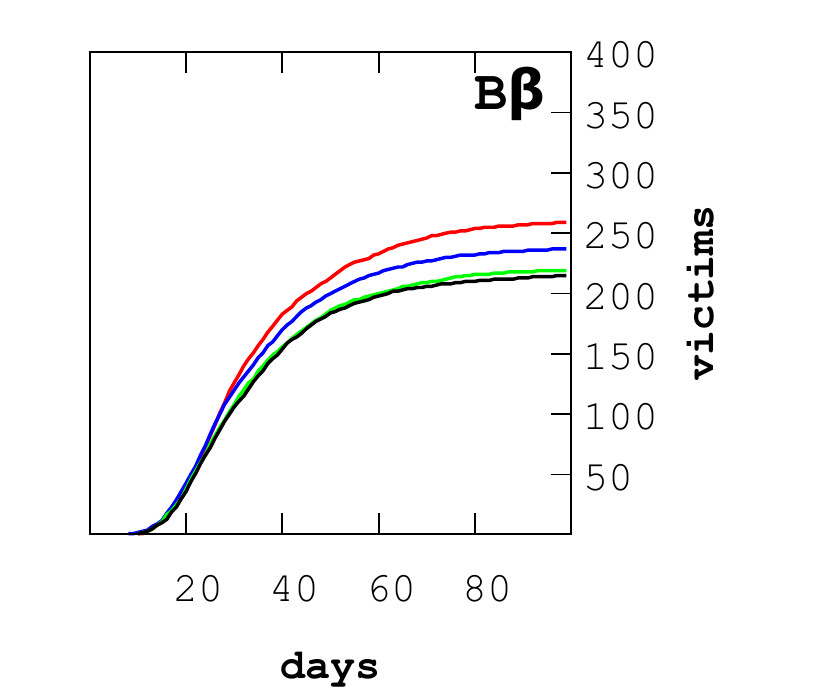}{0.35\textwidth} \\
      \hspace{-1em}(a) & 
      \hspace{-6em}(b) & 
      \hspace{-7em}(c) & 
      \hspace{-9em}(d) 
    \end{tabular}

    \vspace{-1em}

  \end{center}
  \caption{\capstyle Results for$F=0.05$ ($5\%$ of the population are
    vaccinated) and mode \textbf{connected}. See the caption of
    Figure~\ref{pop_05} for details.}
  \label{conn_05}
\end{figure*}

The second series of simulations assumes a limited capacity to
administer the vaccine, so that at most $p$ people can be vaccinated
on any given day. The total amount of vaccine available in the given
period is still assumed to be sufficient to vaccinate a fraction $F$
of the population (here $F=0.2$), so the daily vaccination capacity is
$p=F/T$ (expressed as the fraction of the population that is
vaccinated each day).

In this case the controlled variable is a list $\psi[t]$,
$t=1,\ldots,T$, which determines, for each day, the fraction of daily
vaccination capacity that is administered to the elderly. That is, on
day $t$, $p\cdot{\psi[t]}$ elderly people and $p\cdot{(1-\psi[t])}$
young people are vaccinated (both these values are expressed as a
fraction of the total population). As in the previous case, the
elderly to be vaccinated are chosen at random with uniform
probability, while the young are chosen according to the selected
mode.

For $T=100$, $\psi$ is a list of $100$ values, so it clearly
unfeasible and uninformative to sample it and present all the possible
sequences in a graph. Instead, we use an optimization algorithm to
determine the sequence $\psi[t]$ that minimizes the number of victims
during the vaccination period. The optimization is done using a
genetic algorithm whose details are given in Appendix~\ref{genetic}.

\section{Results}

\subsection{One-shot vaccination}
In this section, the results are presented in the hypothesis that the
delivery capacity is infinite, the only limitation being the amount of
vaccine available. The three parameters that affect the results are
those in Section~\ref{methods}. On day $0$, a number of young and
elderly people are vaccinated and from that point on we track the
number of fatality for $T=100$ days. The infection begins at $t=0$
with one randomly chosen young person infected (the infection never
starts with the elderly).  Figures~\ref{pop_05} and \ref{conn_05} show
the data in the hypothesis that the amount of vaccine available is
enough to vaccinate 5\% of the population. In Figure~\ref{pop_05}, the
young have been vaccinated using the popular mode, while in
Figure~\ref{conn_05} the connected mode was used.
\begin{figure*}[thbp]
  \begin{center}
    \begin{tabular}{cccc}
      \multicolumn{4}{c}{Mode: \textbf{popular}; vaccination level: \textbf{10\%} ($F=0.1$)} \\
      \hspace{-2em}\HInsert{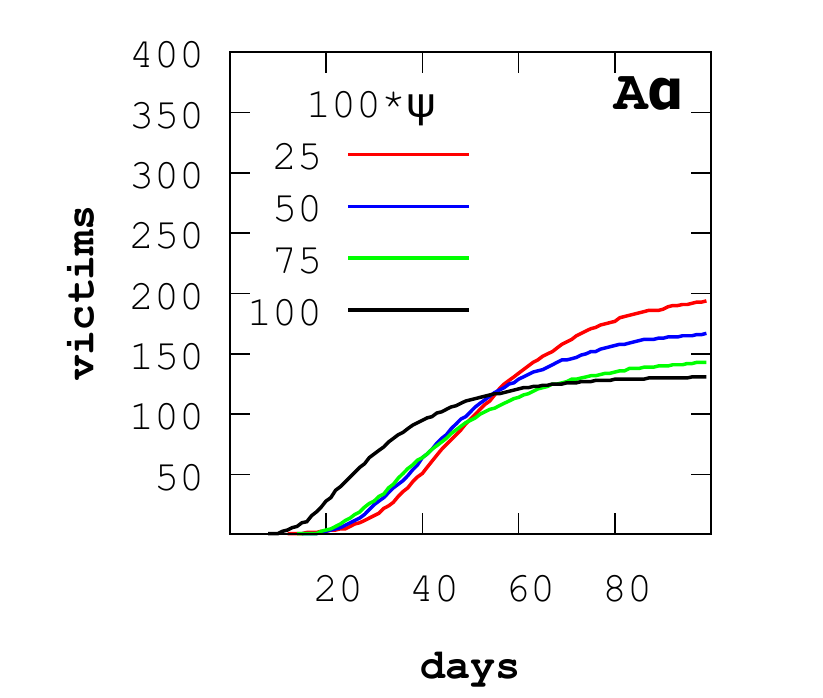}{0.35\textwidth} &
      \hspace{-5.5em}\HInsert{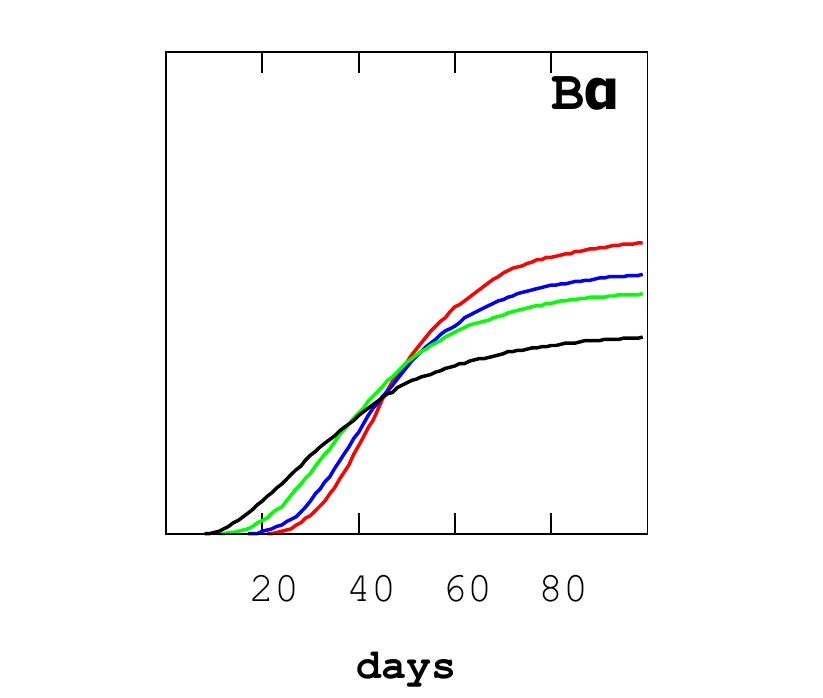}{0.35\textwidth} &
      \hspace{-6.5em}\HInsert{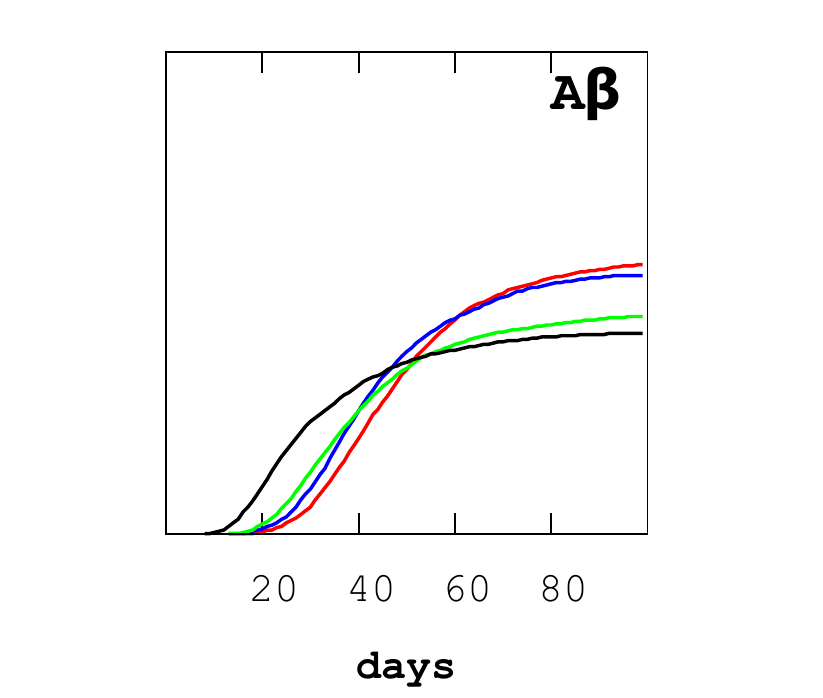}{0.35\textwidth} &
      \hspace{-5.3em}\HInsert{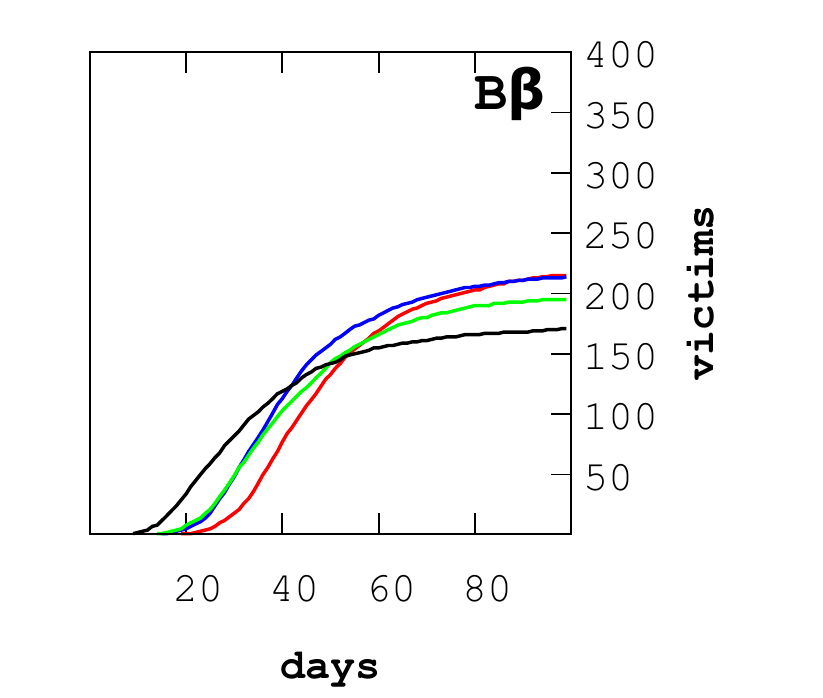}{0.35\textwidth} \\
      \hspace{-1em}(a) & 
      \hspace{-6em}(b) & 
      \hspace{-7em}(c) & 
      \hspace{-9em}(d) 
    \end{tabular}

    \vspace{-1em}

  \end{center}
  \caption{\capstyle Results for$F=0.1$ ($10\%$ of the population are
    vaccinated) and mode \textbf{popular}. See the caption of
    Figure~\ref{pop_05} for details.}
  \label{pop_10}
\end{figure*}

Each series has four graphs, one for each of the graph types of
Table~\ref{epivals}, and the four curves in each graph are the
victims for different values of $\psi$, the percentage of doses that are
given to the elderly ($\psi=0.25,0.5,0.75,1$). In almost all cases, at
the end of the $100$ days, the result is the expected: the number of
victims decreases as the fraction of vaccine administered to the
elderly increases. The only exception is Figure~\ref{conn_05}.a, in
which the differences are not statistically significant.

The smallest number of victims is achieved in
Figure~\ref{conn_05}.a. This is the graph $A\alpha$, with little
contact among the elderly and little contact between the elderly and
the young. The value is lower in Figure~\ref{conn_05}.a, the connected
mode, suggesting that if the vulnerable population is relatively
isolated, vaccinating the (relatively few) contacts is a good
strategy.
\begin{figure*}[thbp]
  \begin{center}
    \begin{tabular}{cccc}
      \multicolumn{4}{c}{Mode: \textbf{connected}; vaccination level: \textbf{10\%} ($F=0.1$)} \\
      \hspace{-2em}\HInsert{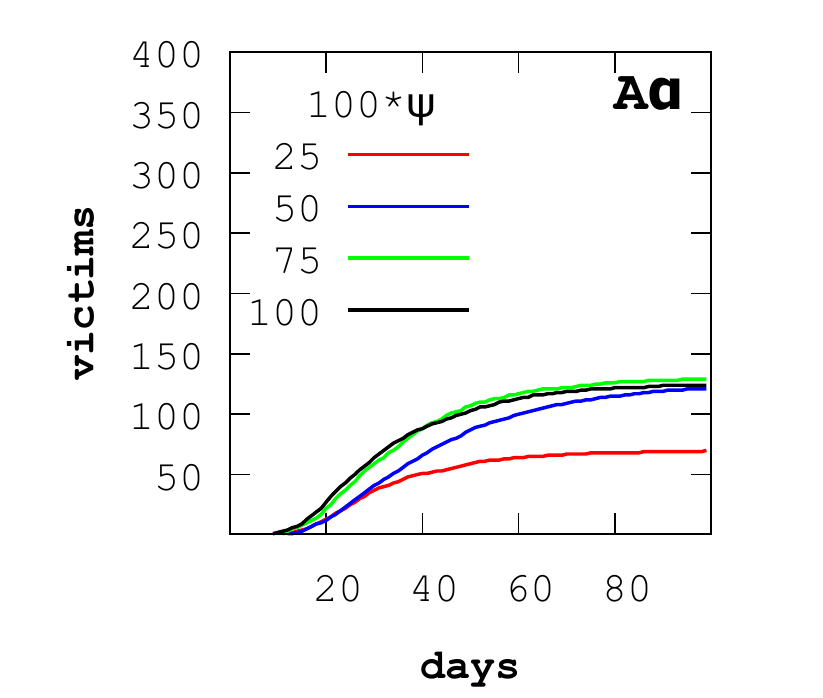}{0.35\textwidth} &
      \hspace{-5.5em}\HInsert{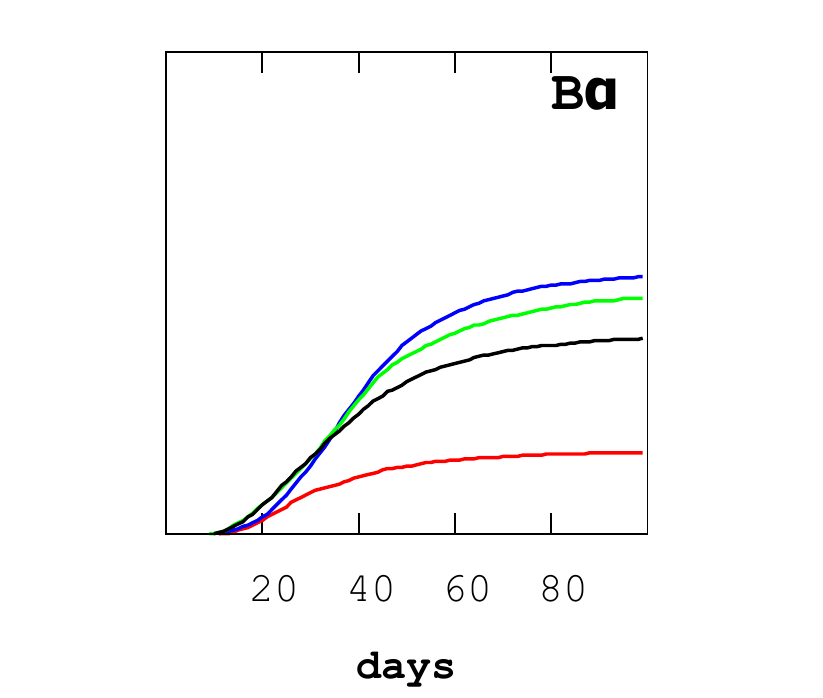}{0.35\textwidth} &
      \hspace{-6.5em}\HInsert{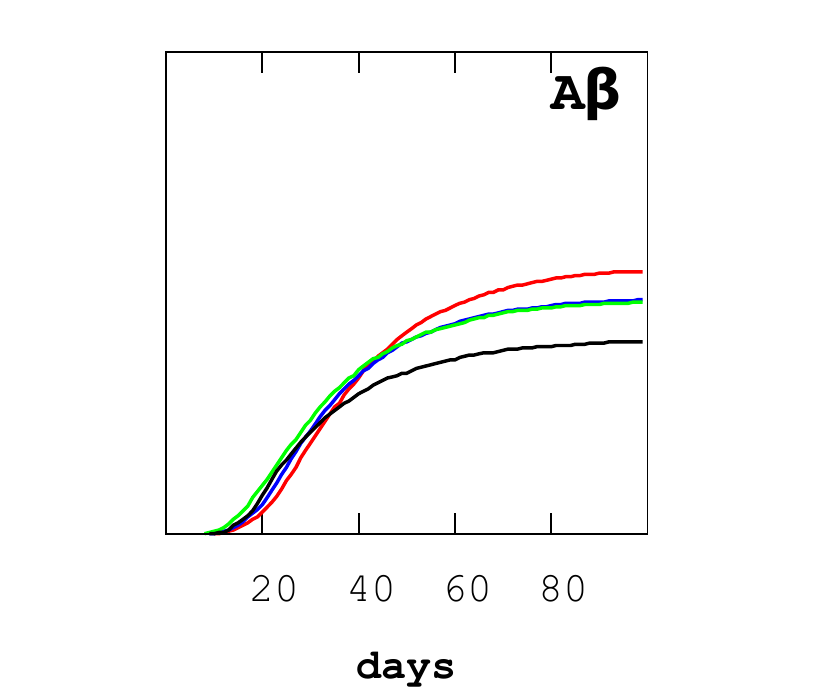}{0.35\textwidth} &
      \hspace{-5.3em}\HInsert{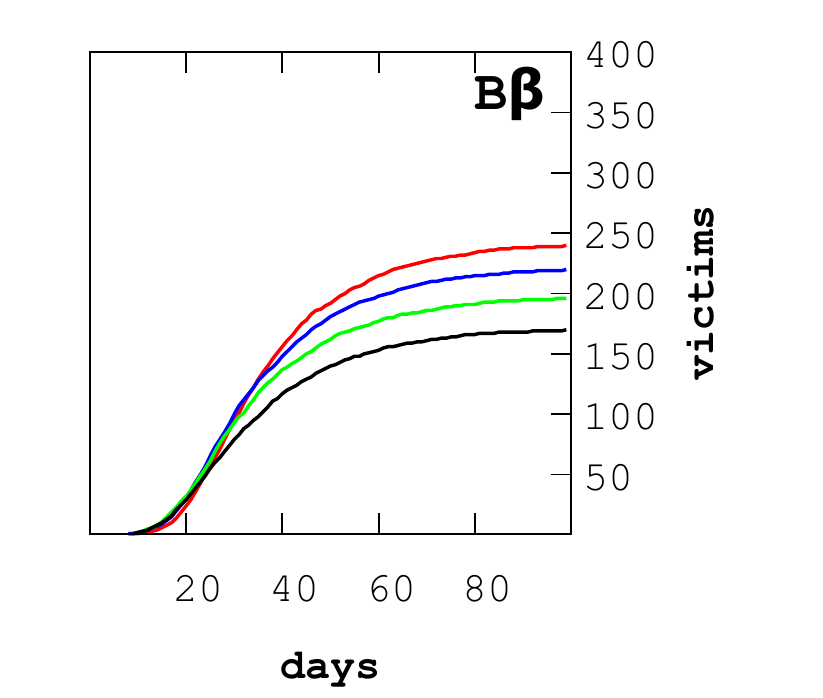}{0.35\textwidth} \\
      \hspace{-1em}(a) & 
      \hspace{-6em}(b) & 
      \hspace{-7em}(c) & 
      \hspace{-9em}(d) 
    \end{tabular}

    \vspace{-1em}

  \end{center}
  \caption{\capstyle Results for$F=0.1$ ($10\%$ of the population are
    vaccinated) and mode \textbf{connected}. See the caption of
    Figure~\ref{pop_05} for details.}
  \label{conn_10}
\end{figure*}

A peculiar phenomenon, which we call \emph{risk inversion} takes place
in the popular mode, especially in the A$\beta$ and B$\beta$ graphs:
if all doses are given to the elderly ($\psi=10$) the number of victim
in the first 20-30 days is higher than if some of the doses are given
to young people. In this case, vaccinating only the elderly does offer
some protection to the most vulnerable (albeit an incomplete one:
there are, in the case $F=0.05$, not enough doses to vaccinate all the
elderly), but leaves the more heavily connected young completely
exposed, leading to a rapid expansion of the epidemic. One the one
hand this leads, in the $\beta$ scenarios in which young and old are
more connected, to more victims among the unprotected elderly and, on
the other hand, to relatively many victims among the young. On a
longer time span, the lack of protection of the elderly for low $\psi$
compensates the effects of slow spreading, and the victims increase
when $\psi$ is small.

\begin{figure*}[thbp]
  \begin{center}
    \begin{tabular}{cccc}
      \multicolumn{4}{c}{Mode: \textbf{popular}; vaccination level: \textbf{20\%} ($F=0.2$)} \\
      \hspace{-2em}\HInsert{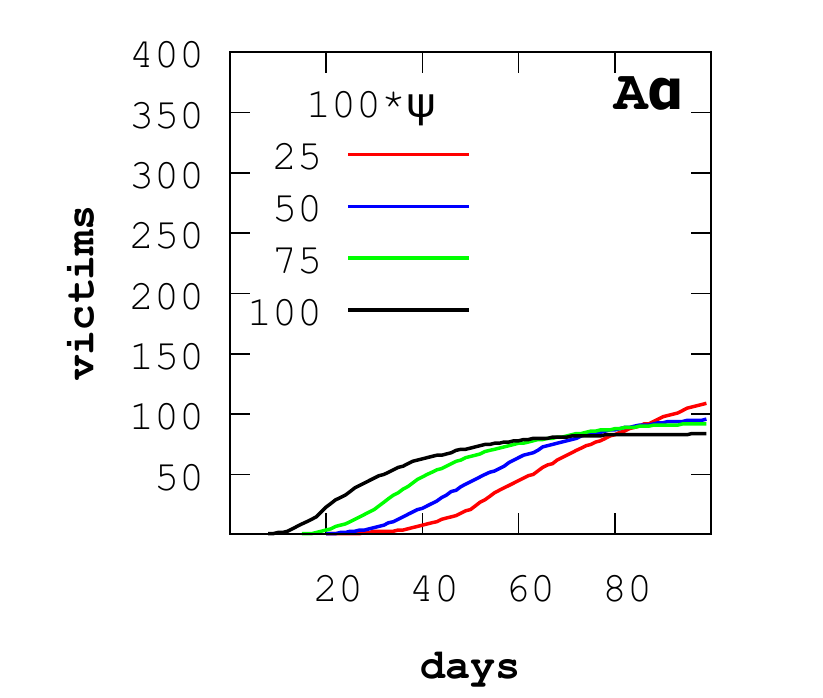}{0.35\textwidth} &
      \hspace{-5.5em}\HInsert{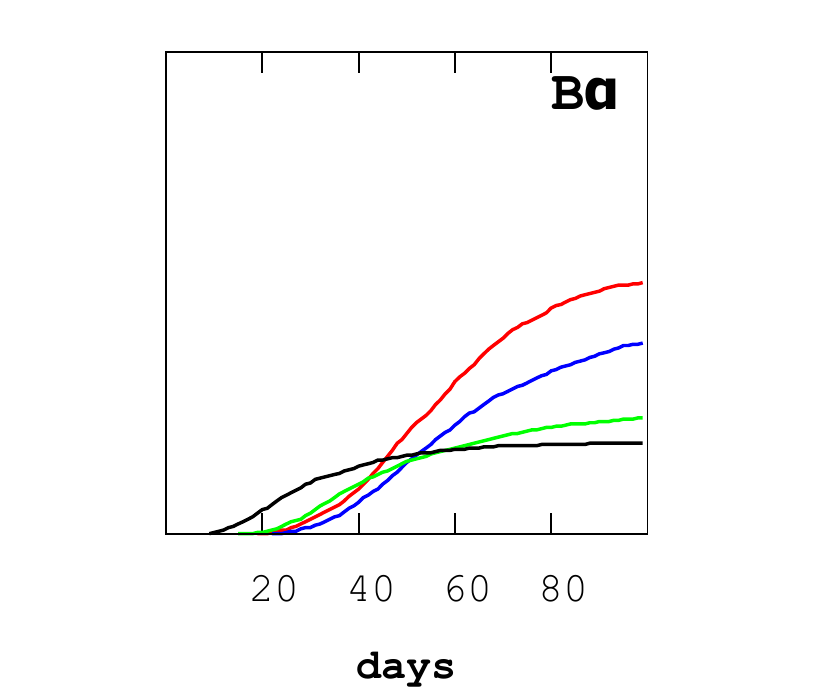}{0.35\textwidth} &
      \hspace{-6.5em}\HInsert{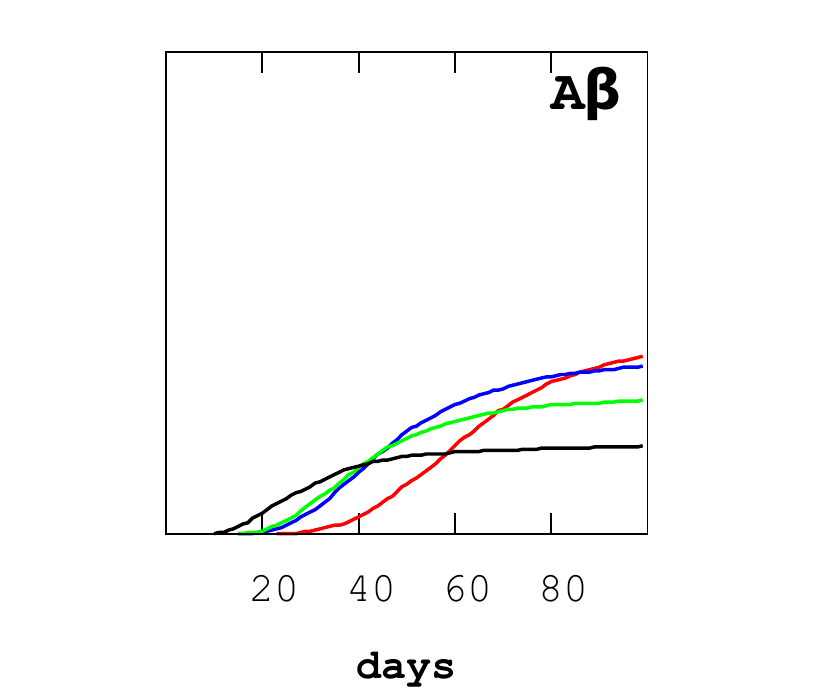}{0.35\textwidth} &
      \hspace{-5.3em}\HInsert{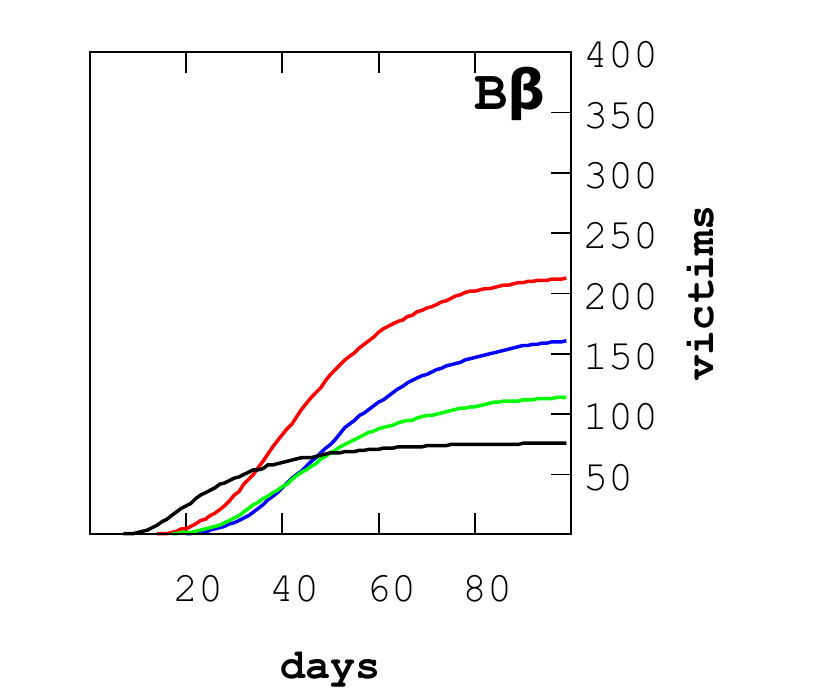}{0.35\textwidth} \\
      \hspace{-1em}(a) & 
      \hspace{-6em}(b) & 
      \hspace{-7em}(c) & 
      \hspace{-9em}(d) 
    \end{tabular}

    \vspace{-1em}

  \end{center}
  \caption{\capstyle Results for$F=0.2$ ($20\%$ of the population are
    vaccinated) and mode \textbf{popular}. See the caption of
    Figure~\ref{pop_05} for details.}
  \label{pop_20}
\end{figure*}

\begin{figure*}[thbp]
  \begin{center}
    \begin{tabular}{cccc}
      \multicolumn{4}{c}{Mode: \textbf{connected}; vaccination level: \textbf{20\%} ($F=0.2$)} \\
      \hspace{-2em}\HInsert{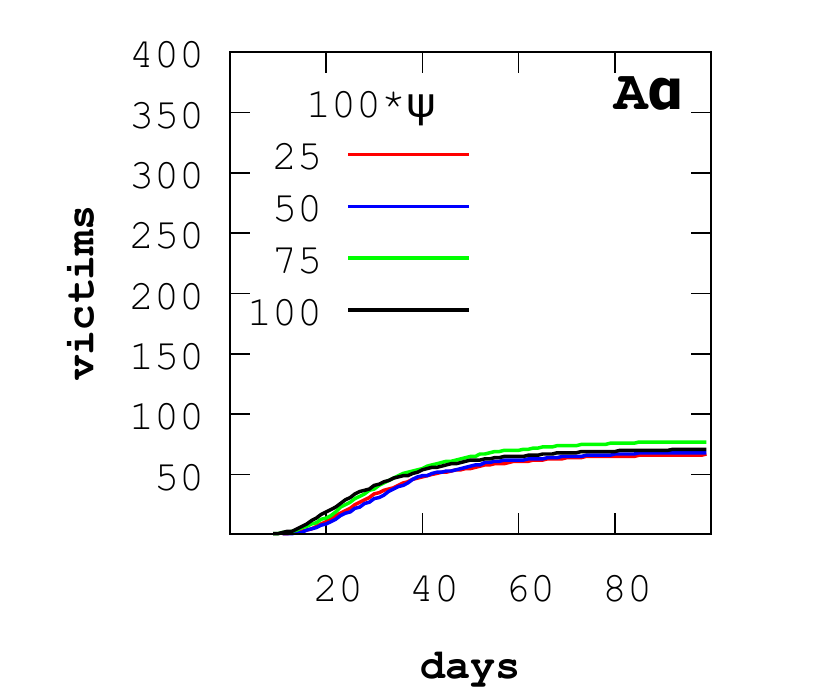}{0.35\textwidth} &
      \hspace{-5.5em}\HInsert{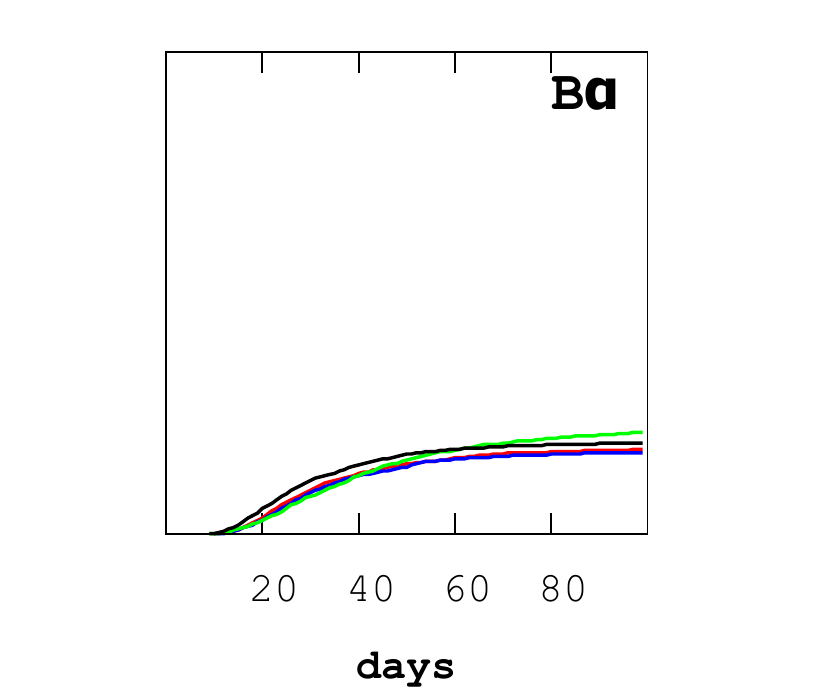}{0.35\textwidth} &
      \hspace{-6.5em}\HInsert{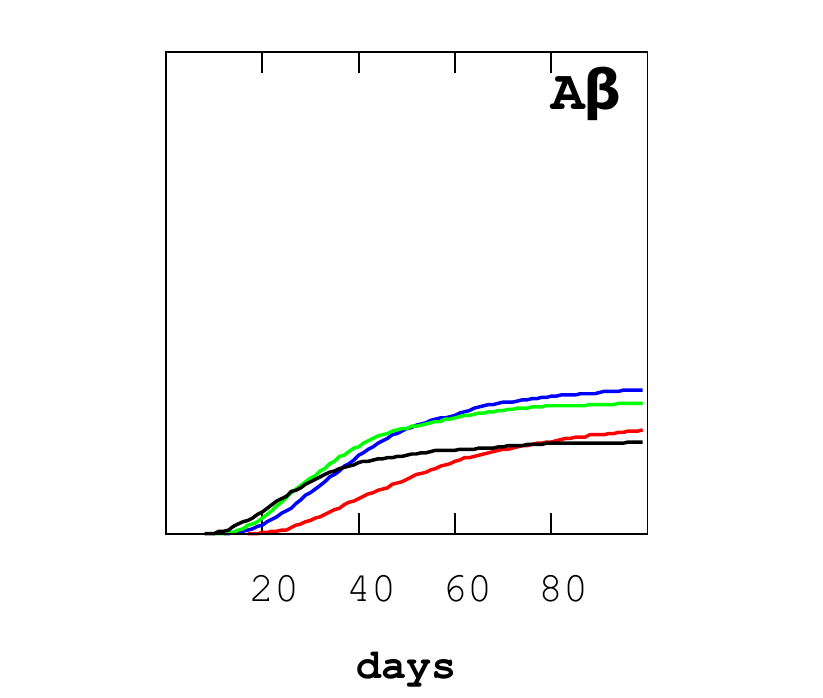}{0.35\textwidth} &
      \hspace{-5.3em}\HInsert{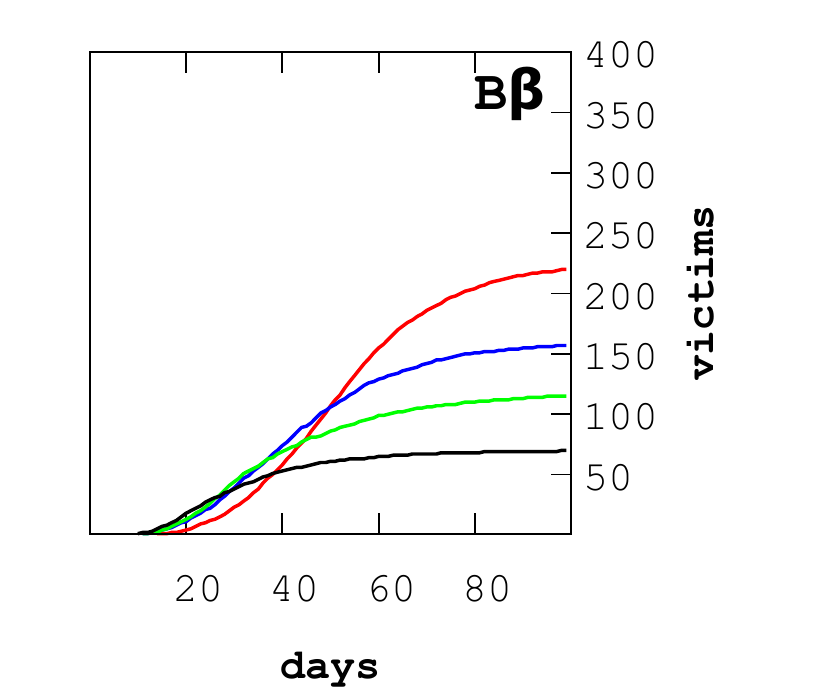}{0.35\textwidth} \\
      \hspace{-1em}(a) & 
      \hspace{-6em}(b) & 
      \hspace{-7em}(c) & 
      \hspace{-9em}(d) 
    \end{tabular}

    \vspace{-1em}

  \end{center}
  \caption{\capstyle Results for$F=0.2$ ($20\%$ of the population are
    vaccinated) and mode \textbf{connected}. See the caption of
    Figure~\ref{pop_05} for details.}
  \label{conn_20}
\end{figure*}

Figures~\ref{pop_10} and \ref{conn_10} show the results when 10\% of
the population is vaccinated, while Figures~\ref{pop_20} and
\ref{conn_20} are relative to a vaccination level of 20\%.
Qualitatively, most of the results are similar. It is noteworthy the
sharp decrease in mortality in Figure~\ref{conn_10}.a with respect
to Figure~\ref{conn_05}.a, which suggests that relative social
distancing is key to obtain the best results in a situation of
scarcity of vaccine. The risk inversion is present here as in the
previous cases.

One striking result is that of Figure~\ref{conn_10}.a and
\ref{conn_10}.b, in which the mortality behaves opposite as in
other cases: giving only 25\% of the doses to the elderly results in a
significant decrease in mortality. These figures correspond to the
\emph{connected} mode of the $\alpha$ graphs, with scarce connection
between the elderly and the young. In this situation of high isolation
and relatively high availability of doses, using 75\% of the vaccine
for the connected young effectively isolates the elderly while
reducing the victims among the young. Compare this with
Figures~\ref{conn_20}.a and \ref{conn_20}.b. In this case, the number
of doses is sufficient to provide a better isolation to the vulnerable
group, so that in the $\alpha$ graphs the number of victims is low
regardless the value of $\psi$. In the $\beta$ graph of
Figure~\ref{conn_20}.b, the stronger connection between elderly and
young balances out the higher number of doses available, and the
phenomenon of Figure~\ref{conn_10}.a is repeated. The convenience of
targeting the \dqt{connection points} (the young people that have
connections with the vulnerable population) depends on the amount of
vaccine available and on the density of the relations between elderly
and young.

The risk inversion in the early days of vaccination is present in all
cases, although not with the same prominence. This phenomenon suggests
that if the vaccine is available in batches, the vaccination policy
might be different for different batches: a first vaccination in part
to the young to limit the speed of the infection and, if a second
batch is available within 20 or 30 days, a more massive vaccination to
the elderly to protect them.

\subsection{Incremental vaccination}
\begin{figure*}[htbp]
  \begin{center}
    \begin{tabular}{cccc}
      \multicolumn{4}{c}{Mode: \textbf{popular}} \\
      \hspace{-2em}\HInsert{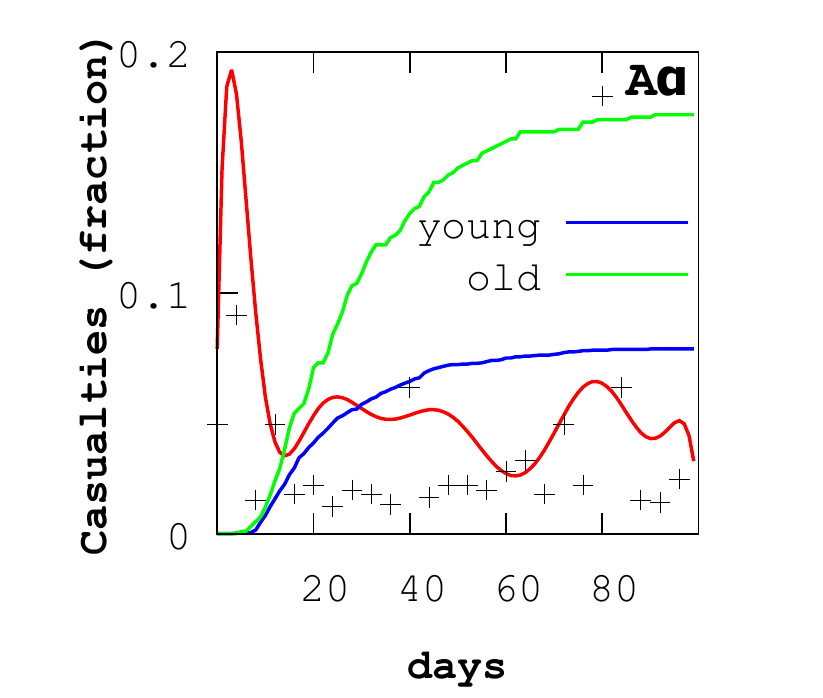}{0.35\textwidth} &
      \hspace{-5.5em}\HInsert{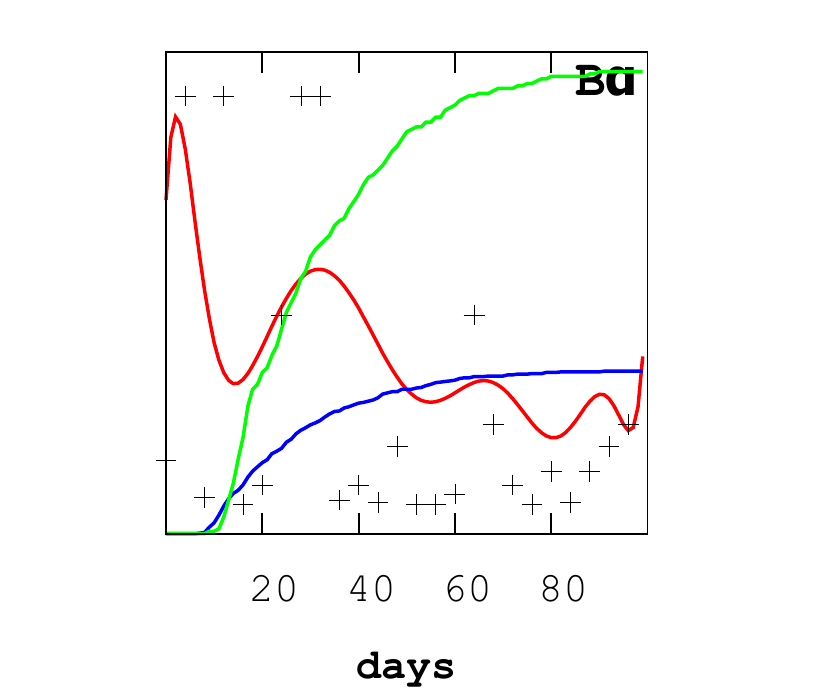}{0.35\textwidth} &
      \hspace{-6.5em}\HInsert{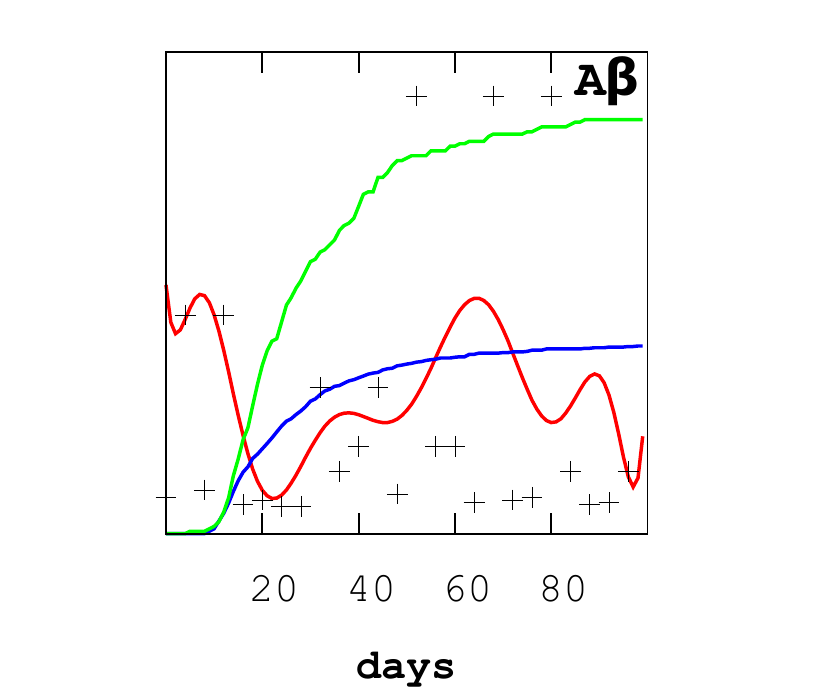}{0.35\textwidth} &
      \hspace{-5.3em}\HInsert{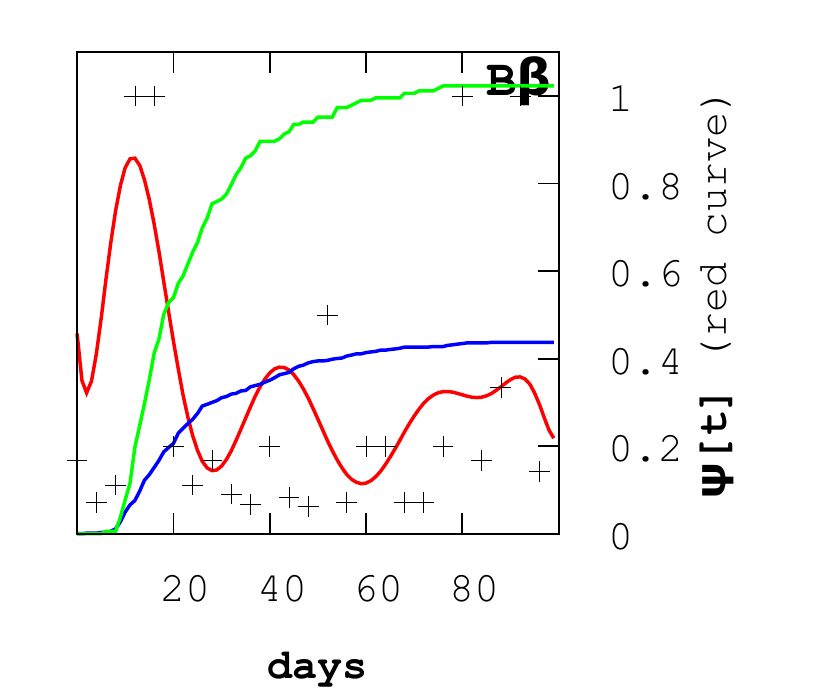}{0.35\textwidth} \\
      \hspace{-1em}(a) & 
      \hspace{-6em}(b) & 
      \hspace{-7em}(c) & 
      \hspace{-9em}(d) 
    \end{tabular}
  \end{center}
  \caption{\capstyle Optimal vaccination strategy and the resulting
    mortality. The black crosses are the output of the optimizer; the
    red line is a polynomial approximation (degree=10) that highlights
    the general trends of the solution, the blue and green lines are the
    fraction of the young and elderly that have died, respectively,
    each one expressed as a fraction of the respective population. The
    vaccination of the young is done using the popular mode: the
    young people most connected are vaccinated first.}
  \label{inc_pop}
\end{figure*}

\begin{figure*}[htbp]
  \begin{center}
    \begin{tabular}{cccc}
      \multicolumn{4}{c}{Mode: \textbf{connected}} \\
      \hspace{-2em}\HInsert{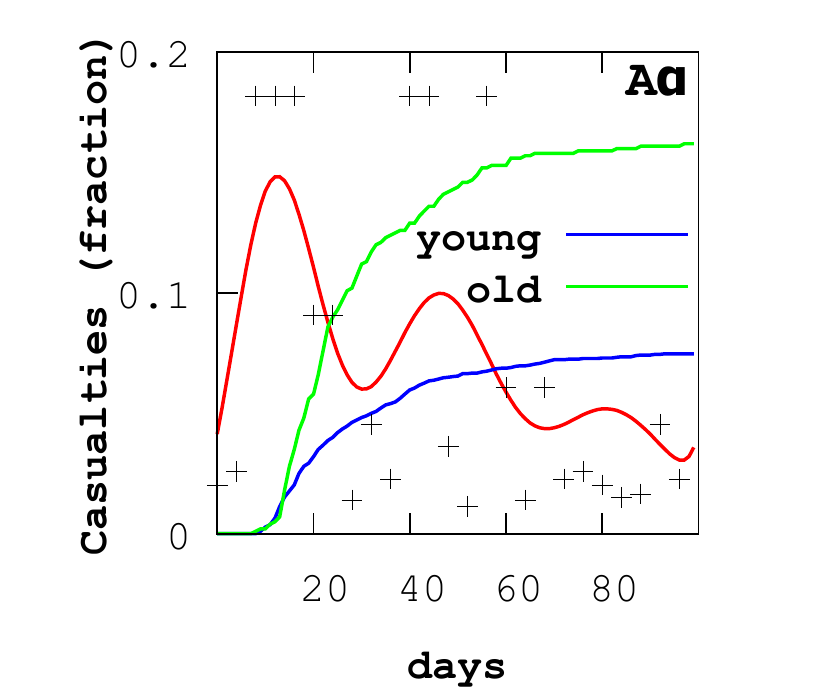}{0.35\textwidth} &
      \hspace{-5.5em}\HInsert{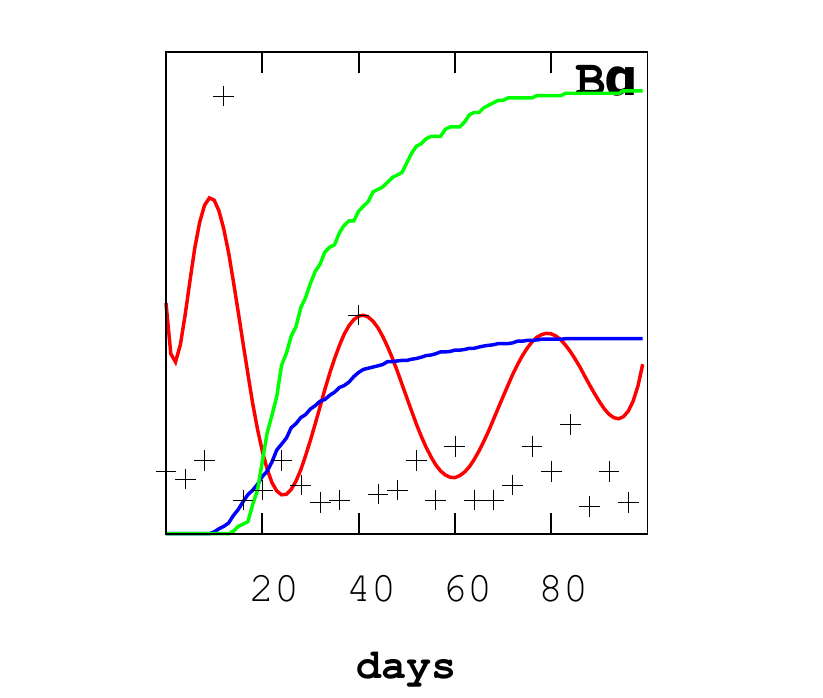}{0.35\textwidth} &
      \hspace{-6.5em}\HInsert{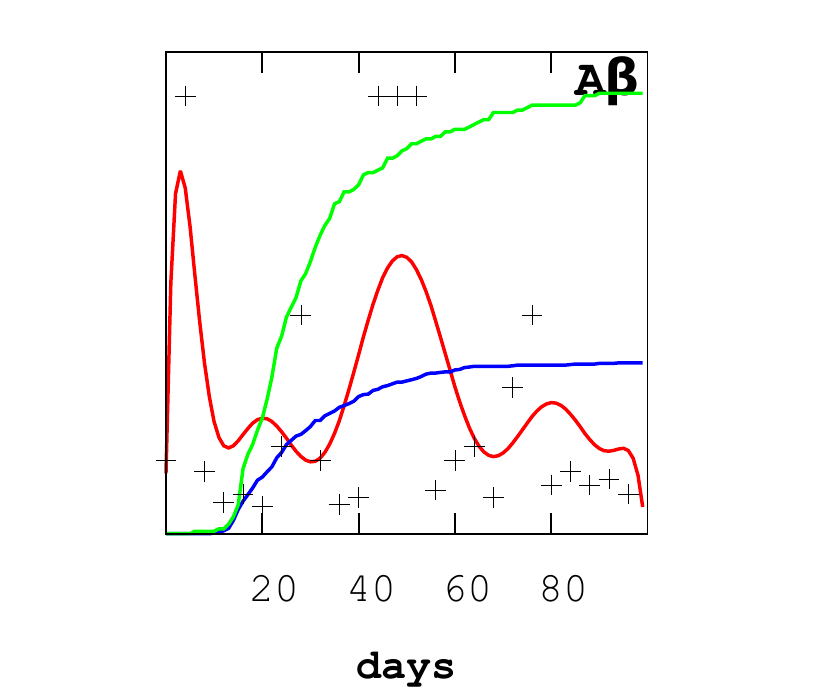}{0.35\textwidth} &
      \hspace{-5.3em}\HInsert{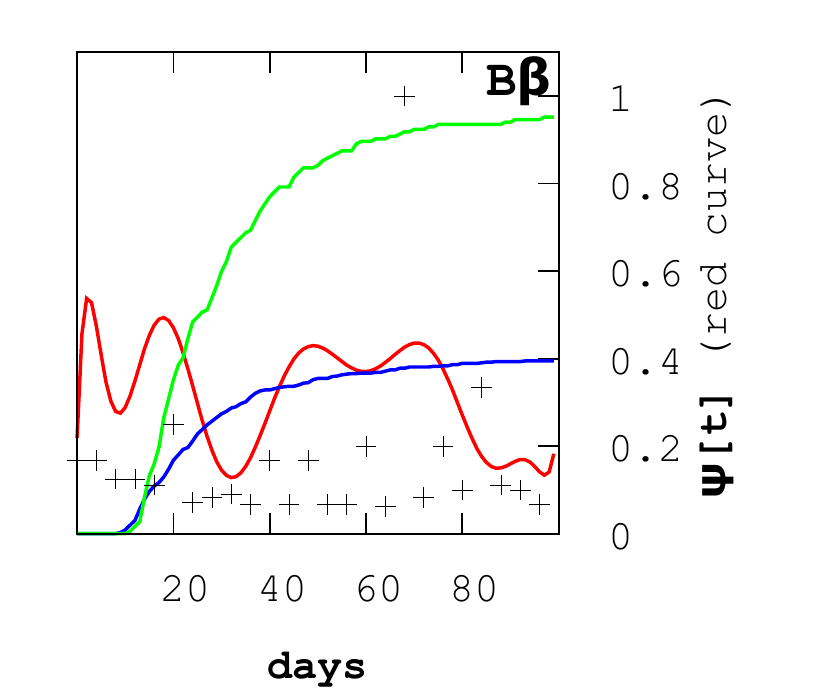}{0.35\textwidth} \\
      \hspace{-1em}(a) & 
      \hspace{-6em}(b) & 
      \hspace{-7em}(c) & 
      \hspace{-9em}(d) 
    \end{tabular}
  \end{center}
  \caption{\capstyle Optimal vaccination strategy and the resulting
    mortality. The black crosses are the output of the optimizer; the
    red line is a polynomial approximation (degree=10) that highlights
    the general trends of the solution, the blue and green lines are the
    fraction of the young and elderly that have died, respectively,
    each one expressed as a fraction of the respective population. The
    vaccination of the young is done using the connected mode: the
    young people most connected to the elderly are vaccinated first.}
  \label{inc_conn}
\end{figure*}

Figures~\ref{inc_pop} and \ref{inc_conn} show the optimal vaccination
schedule as determined by the optimization algorithm. The vaccination
schedule is assumed to extend over a period of 100 days, and the total
number of doses is assumed to be sufficient to vaccinate 20\% of the
population. The results were obtained using 500 generations of 800
genes each (see Appendix~\ref{genetic}). The results of the
best genes of several generations were checked to verify that the solutions were
stable. In the figures, the black crosses mark the output of the
optimizer, that is, they represent the optimal fraction of vaccine to
be administered to the elderly each day. The red line is a polynomial
least squares approximation of these data using a polynomial of degree
10, which allows us to detect the general trends of the solution
without the abrupt changes: our considerations will be based on these
curves. The blue and green lines are the victims among the young and old
people respectively, expressed as a fraction of the respective
population. In Figure~\ref{inc_pop} the vaccination of the young is
done using the popular mode, while in Figure~\ref{inc_conn} the
connected mode was used. The four graphs in each figure are relative
to the four graph types of Table~\ref{epivals}.

One common feature of all the strategies is the high initial fraction
of vaccine administered to the elderly followed by an increase in
vaccine given to the young when the number of casualties begins to
grow. The peaks are more pronounced in the popular mode: in this case,
after an initial short vaccination period to protect part of the
vulnerable population, the focus moves to the young in order to stop
the spread of infection by cutting the social hubs through
vaccination. The situation is less clear in the $\beta$ graphs (with
many connections between the two groups). In this case, especially in
the A$\beta$ graph, the optimal strategy seems to involve a more
balanced approach, with alternating focus on the elderly and the
young.

For the connected mode, in the $\alpha$ graphs, the peak of
vaccination of the elderly is wider. Given the low value of
$\bar{d}_{ey}$, relatively few vaccinations of the young are sufficient
to block the spread of the infection to the vulnerable population, so
the initial vaccination effort can concentrate on the vulnerable for a
longer period of time. In the A$\beta$ graph, with few
elderly-to-elderly relations and strong connections with the young,
the focus shifts on a much earlier vaccination of the young.

\section{Conclusions}
We have developed a model that, with respect to the standard SIR
model, allows us to take into account the variability of the social
relations of people in a social scenario. Given the existence of
social groups especially vulnerable to the disease (herein referred to
as \dqt{the elderly}), two factors appear crucial when defining a
vaccination strategy: the social relations among the elderly, and
the connections between the elderly and the young.

The most general result that we can draw from the model is that a
prudent vaccination strategy should balance two requirements: protect
the elderly, and vaccinate enough young people (with intense social
relations) to slow down the spread of the infection.

The exact nature of this balance depends on the social connections and
on who, among the young people, is vaccinated.  Overall, the most
effective strategy seems to be to give priority, among the young, to
those who are more heavily connected to the vulnerable people
(caregivers, cleaning personnel, etc.). If the amount of available
vaccine is not too scarce (at least 10\% of the population in our
simulations), the results are good even if only 25\% of the doses is
administered to the vulnerable. If additional measures are taken to
keep the vulnerable isolated, the 25\% solution is actually the one
that gives the best results. If the elderly are not isolated, then the
strategy of splitting the vaccine between them and the young has
positive results in a short (20-30 days) period but, in a longer
period, it is necessary to move to a heavier vaccination to the
vulnerable.

A more detailed strategy can be devised if we take into account a
limited delivery capacity and the distribution of the vaccination over
the whole duration of the epidemic. In this case, the optimal strategy
seems to begin with a higher fraction of elderly vaccinated then, when
the number of victims begins to grow, with a more intense vaccination
of young people to keep the spread under control. The strategy that
follows this first phase is less clear, but it appear to always
consist in an oscillation with periods of more intense vaccination of
the elderly alternated with periods of more intense vaccination of the
young, all with a general tendency to increase the fraction of vaccine
administered to the young.

A few words of caution are in order. All these results were obtained
through a mathematical model in which, necessarily, many of the
details of the actual situation are abstracted. While the general
conclusions that we derived have undoubtedly a validity, applying them
in the field requires extreme caution and a deeper analysis of the
evolution of the epidemic.

Also, our model has assumed that the \dqt{most popular} and \dqt{most
  connected} people can be identified and summoned for
vaccination. This hypothesis might be reasonably realistic for the
connected mode, in which the focus can often be identified by
profession than with the more elusive popular mode. The distribution
of the number of neighbors of the node one arrives to when crossing a
random edge \cite{newman:05} may help implementing this mode.

Finally, we have ignored here all the problems of public perception,
and the media backlash that may be caused by a strategy that, rational
as it might be, does not follow what the people (or, more importantly,
the media) perceive as \dqt{common sense}.

\appendix

\section{Optimization of the vaccination schedule}
\label{genetic}
Mathematically, our problem is the following. We have a
multidimensional variable, the values $\psi[t]\in[0,1]$,
$t=1,\ldots,T$ ($T=100$ in our case), and a function $G(\psi)$ that we
want to minimize. In this case, $G(\psi)$ is the number of fatalities
for the infection when $\psi$ is used as a vaccination schedule
(fraction of the vaccination that is given to the elderly each
day). The value of $G$ is not a trivial function of $\psi$. In
particular, we cannot assume any of the standard properties
(continuity, derivability, analiticity, etc.) that most optimization
algorithms assume. Because of the nature of the function, a genetic
algorithm \cite{mitchell:98} is a reasonable option to solve this optimization
problem.

We transform the problem into a discrete one by restricting $\psi[t]$ to
values of the type
\begin{equation}
  \label{fracgene}
  \psi[t] = \frac{1}{n[t]+1}
\end{equation}
with $n[t]\in[0,15]$. Each $n[t]$ can be represented as a four-bit
number, so the sequence $n[1],\ldots,n[T]$ that determines the
solution can be represented as an integer of $4\cdot{T}$ bits. This
is the \dqt{gene} of our algorithm.

A generation is composed of a collection of $G$ genes $\gamma_i$,
$i=1,\ldots,G$. Given a gene, we derive from it the vaccination
schedule using (\ref{fracgene}) and execute the simulation. This will
result in a score $s_i$. Standard genetic algorithms are function
maximizers, so we consider as score the inverse of the number of
fatalities at time $T$ ($s_i=-V_{\gamma_i}(T)$).
\begin{figure}[tbhp]
  \begin{center}
    {\tt
    \setlength{\unitlength}{1em}
    \begin{picture}(20,12)(0,0)
      \put(4,1){\line(0,1){11}}
      \put(11,1){\line(0,1){11}}
      \put(4,0.8){\makebox(0,0)[t]{a}}
      \put(11,0.8){\makebox(0,0)[t]{b}}
      \put(0,2){
        \thicklines
        \multiput(0,0)(0,1){2}{\line(1,0){14}}
        \multiput(0,0)(14,0){2}{\line(0,1){1}}
        \thinlines
        \put(2,0.5){\makebox(0,0){$B_1$}}
        \put(7.5,0.5){\makebox(0,0){$A_2$}}
        \put(12.5,0.5){\makebox(0,0){$B_3$}}
        \put(16,0.5){\makebox(0,0)[l]{offspring 2}}
      }
      \put(0,4){
        \thicklines
        \multiput(0,0)(0,1){2}{\line(1,0){14}}
        \multiput(0,0)(14,0){2}{\line(0,1){1}}
        \thinlines
        \put(2,0.5){\makebox(0,0){$A_1$}}
        \put(7.5,0.5){\makebox(0,0){$B_2$}}
        \put(12.5,0.5){\makebox(0,0){$A_3$}}
        \put(16,0.5){\makebox(0,0)[l]{offspring 1}}
      }
      \put(0,7){
        \thicklines
        \multiput(0,0)(0,1){2}{\line(1,0){14}}
        \multiput(0,0)(14,0){2}{\line(0,1){1}}
        \thinlines
        \put(2,0.5){\makebox(0,0){$B_1$}}
        \put(7.5,0.5){\makebox(0,0){$B_2$}}
        \put(12.5,0.5){\makebox(0,0){$B_3$}}
        \put(16,0.5){\makebox(0,0)[l]{parent B}}
      }
      \put(0,9){
        \thicklines
        \multiput(0,0)(0,1){2}{\line(1,0){14}}
        \multiput(0,0)(14,0){2}{\line(0,1){1}}
        \thinlines
        \put(2,0.5){\makebox(0,0){$A_1$}}
        \put(7.5,0.5){\makebox(0,0){$A_2$}}
        \put(12.5,0.5){\makebox(0,0){$A_3$}}
        \put(16,0.5){\makebox(0,0)[l]{parent A}}
      }
    \end{picture}
    }
  \end{center}
  \caption{\capstyle Double cut for the generation of offspring. Two parents
    generate two offspring, mixing the bits of their genes as
    represented.}
  \label{offsprings}
\end{figure}
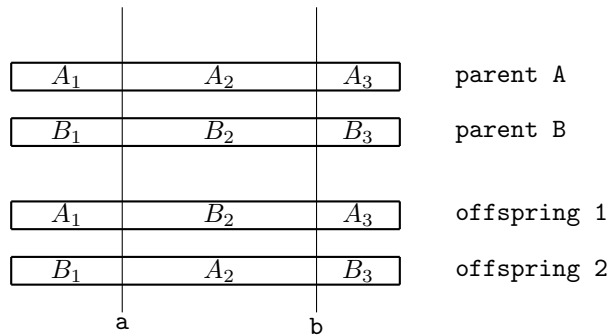
There are several methods to create the following generation of
individuals. Since the performance of the algorithms seems to have
little dependence in the specific method used, we use one of the
simplest, based on the creation of an intermediate \emph{gene
  pool}. The gene pool is a set ${\mathcal{P}}$ of $P$ individuals
possibly replicated (generally $|P|=G$: the pool has the same size
as the generations) such that the number of \dqt{copies} of an
individual in the pool is proportional to its score. An easy algorithm
for generating a pool is the \emph{tournament}: we do $P$ comparisons
of pairs of individuals taken at random from the generation: the
individual with the highest score goes into the pool:

\parbox{\columnwidth}{
  {\tt
    \cb
    \> $P$ $\leftarrow$ $\emptyset$ \\
    \> \cmd{for} k=1 \cmd{to} $P$ \cmd{do} \\
    \> \> i $\leftarrow$ rnd(1,G) \\
    \> \> j $\leftarrow$ rnd(1,G) \\
    \> \> \cmd{if} $s_i\ge{s_j}$ \cmd{then} \\
    \> \> \> ${\mathcal{P}}$ $\leftarrow$ ${\mathcal{P}} \cup \{\gamma_i\}$ \\
    \> \> \cmd{else} \\
    \> \> \> ${\mathcal{P}}$ $\leftarrow$ ${\mathcal{P}} \cup \{\gamma_j\}$ \\
    \> \> \cmd{fi} \\
    \> \cmd{od}
    \ce
  }
}

In order to build the next generation, pairs of genes are taken at
random from the pool (with uniform distribution) and crossed to create
two new individuals that will go into the next generation (this
requires that $G$ be even). We use the method of the \emph{double cut}
to cross the genes. Two values $a,b\in[0,4T]$ are chosen randomly. The
two offspring are then generated as in Figure~\ref{offsprings}, in
which we assume $a<b$. We also define a small mutation probability:
for each new gene, with a (small) probability $p$, we pick a random
bit and flip it.
Note that this method doesn't guarantee that the best individual of a
generation will pass unchanged to the next, so we actually use the
crossing to create $G-2$ individuals to which we add the two best
performers of the previous generation.

\end{document}